\documentclass[sigconf, twocolumn]{acmart}
\pdfoutput=1

\usepackage{tikz}
\usetikzlibrary{shadows.blur}
\usepackage{pgfplots}
\usepgfplotslibrary{groupplots}
\usepackage{algorithm}
\usepackage{algorithmicx}
\usepackage{algpseudocode}
\usepackage{booktabs}
\usepackage{epigraph}

\AtBeginDocument{%
  \providecommand\BibTeX{{%
    \normalfont B\kern-0.5em{\scshape i\kern-0.25em b}\kern-0.8em\TeX}}}

\setcopyright{none}

\acmConference[Woodstock '18]{Woodstock '18: ACM Symposium on Neural
  Gaze Detection}{June 03--05, 2018}{Woodstock, NY}
\acmBooktitle{Woodstock '18: ACM Symposium on Neural Gaze Detection,
  June 03--05, 2018, Woodstock, NY}
\acmPrice{15.00}
\acmISBN{978-1-4503-XXXX-X/18/06}

\renewcommand\footnotetextcopyrightpermission[1]{}
\begin{document}

\title{DB-BERT: a Database Tuning Tool that ``Reads the Manual''}

%%
%% The "author" command and its associated commands are used to define the authors and their affiliations.
% \author{Immanuel Trummer}
% \affiliation{%
%   \institution{Cornell, Database Group}
%   \city{Ithaca}
%   \state{NY}
%   \postcode{14850}
% }
% \email{itrummer@cornell.edu}
\author{Immanuel Trummer}
\affiliation{%
  \institution{Cornell University}
  \city{Ithaca}
  \state{New York}
}
\email{itrummer@cornell.edu}

\begin{abstract}
DB-BERT is a database tuning tool that exploits information gained via natural language analysis of manuals and other relevant text documents. It uses text to identify database system parameters to tune as well as recommended parameter values. DB-BERT applies large, pre-trained language models (specifically, the BERT model) for text analysis. During an initial training phase, it fine-tunes model weights in order to translate natural language hints into recommended settings. At run time, DB-BERT learns to aggregate, adapt, and prioritize hints to achieve optimal performance for a specific database system and benchmark. Both phases are iterative and use reinforcement learning to guide the selection of tuning settings to evaluate (penalizing settings that the database system rejects while rewarding settings that improve performance). In our experiments, we leverage hundreds of text documents about database tuning as input for DB-BERT. We compare DB-BERT against various baselines, considering different benchmarks (TPC-C and TPC-H), metrics (throughput and run time), as well as database systems (Postgres and MySQL). In all cases, DB-BERT finds the best parameter settings among all compared methods. The code of DB-BERT is available online at \url{https://itrummer.github.io/dbbert/}.
\end{abstract}

\maketitle
\pagestyle{plain}
\thispagestyle{empty}

\section{Introduction}

%\epigraph{All human things are subject to decay, and when fate summons, Monarchs must obey}{\textit{Mac Flecknoe \\ John Dryden}}

\begin{epigraphs}
\qitem{{\itshape Give me a user manual, and I'm happy for hours.}}{--- Lennon Parham}
\qitem{{\itshape When all else fails, read the instructions.}}{--- Anonymous}
\end{epigraphs}

Manuals are useful. For instance, before starting to tune a database management system (DBMS), it is recommended to read the associated manual. So far, those words of wisdom only seemed to apply to human database administrators. While it is widely acknowledged that database manuals contain useful information, this knowledge has long been considered inaccessible to machines due to barriers in natural language understanding. We believe that this has changed with recent advances in the field of natural language processing, namely by the introduction of powerful, pre-trained language models based on the Transformer architecture~\cite{Vaswani2017}. We present DB-BERT, a tuning tool, based on the BERT model~\cite{Devlin2019a}, that ``reads'' (i.e., analyzes via natural language tools) the manual and hundreds of text documents with tuning hints in order to find promising settings for database system parameters faster.

The problem of finding optimal values for DBMS parameters (also called ``tuning knobs'') for specific workloads and performance metrics has received significant attention in recent years. DBMS nowadays feature hundreds of parameters~\cite{Pavlo2017}, making it very hard to find optimal settings manually. This motivates computational methods for automated parameter tuning. The dominant approach is currently machine learning~\cite{Aken}, in particular reinforcement learning~\cite{VanAken2021, Li2018, Zhang2019a}. Here, a tuning tool selects value combinations for DBMS parameters to try in a principled manner, guided by the results of benchmark runs for specific settings. However, this approach is expensive (recent work uses hundreds of iterations per tuning session~\cite{VanAken2021}) and works best if guided by input from database experts~\cite{Kanellis2020}, pre-selecting a small set of parameters to tune and reasonable value ranges to consider. Our goal is to substitute such input by information that is gained automatically by analyzing text documents. We call the corresponding problem variant \textit{Natural Language Processing (NLP)-Enhanced Database Tuning}.

\begin{table}[t]
    \centering
    \caption{Example tuning hints with extractions.}
    \begin{tabular}{p{5cm}p{2.5cm}}
        \toprule[1pt]
         \textbf{Text Snippet} & \textbf{Extraction} \\
         \midrule[1pt]
         The default value of \verb|shared_buffer| is set very low ... The recommended value is 25\% of your total machine RAM.~\cite{PERCONA2018} & \verb|shared_buffers| $\quad=0.25\cdot RAM$ \\
         \midrule
         I changed ‘\verb|random_page_cost|’ to 1 and retried the query. This time, PostgreSQL used a Nested Loop and the query finished 50x faster.~\cite{Patibandla2017} & \verb|random_page_cost| $=1$ \\
         \midrule
         On a dedicated database server, you might set the buffer pool size to 80\% of the machine's physical memory size.~\cite{MySQL8manual} & \verb|innodb_buffer_| \verb|pool_size| $=0.8\cdot RAM$ \\
         \bottomrule[1pt]
    \end{tabular}
    \label{tab:examples}
\end{table}

DB-BERT extracts, from text, tuning hints that recommend specific values for specific parameters. Table~\ref{tab:examples} shows examples with sources and the associated, formal representation of each extracted hint. Some of the hints (second example) recommend an absolute value while others (first and third example) recommend relative values. For the latter, translating the hint into a concrete value recommendation requires knowledge of system properties such as the amount of RAM. Some of the hints (first two examples) mention the parameter explicitly while others (last example) refer to it only implicitly. DB-BERT can exploit all of the hints shown in Table~\ref{tab:examples}.

For a given text snippet, DB-BERT uses a fine-tuned version of the BERT Transformer model to solve four tasks. First, it decides whether a text snippet contains hints. Second, it translates hints into formulas such as the ones shown in Table~\ref{tab:examples}. This may entail steps for resolving implicit parameter references as well as relative recommendations. Third, instead of relying on hints completely, DB-BERT may decide to deviate from proposed values within pre-defined ranges. Finally, given potentially conflicting hints from multiple sources, DB-BERT chooses weights for hints, representing their relative importance.

DB-BERT does not rely on tuning hints alone. Instead, it uses tuning hints as guidelines for a tuning approach that is based on reinforcement learning. During a tuning session, DB-BERT iterates until a user-defined optimization time budget runs out. In each iteration, DB-BERT selects one or multiple DBMS configurations (i.e., parameter settings) to try out. DB-BERT translates the performance observed during those runs (on user-defined benchmarks) into a reward value. This reward value is used to guide the selection of configurations in future iterations, using the Double Deep Q-Networks~\cite{VanHasselt2016} reinforcement learning algorithm. To apply this algorithm, we formulate database tuning as a Markov Decision Process (MDP) with discrete states and actions. We represent treatment for each hint as a sequence of decisions, determining the hint type (e.g., relative versus absolute values) as well as the hint weight. To leverage NLP for those decisions, we associate each decision option with a text label. This allows DB-BERT to compare hint text and decision label using the BERT Transformer.

We train DB-BERT in a system and benchmark independent manner, before applying it for specific tuning tasks. In principle, we could use manually annotated tuning documents for training (assigning a high reward for hint translations that are consistent with annotations). However, generating such data requires expert knowledge and is hard to crowdsource (compared to cases where labeling requires only commonsense knowledge~\cite{Gatt2017}). Instead, we exploit the database system itself for (noisy) feedback. We assume that tuning hints, if correctly translated, tend to recommend admissible values that do not to dramatically decrease performance. Hence, we train DB-BERT by assigning rewards for hint translations that result in admissible parameter settings (i.e., the DBMS accepts the setting). On the other side, we assign penalties for translations that result in inadmissible parameter settings (i.e., the DBMS rejects the setting) or settings that decrease performance significantly for a simple example workload. The result of training is a model (i.e., weights for around 110~million parameters of the fine-tuned BERT model) that can be used as starting point for tuning other database systems on other benchmarks.

We are only aware of one recent vision paper, aimed at leveraging text documents for database tuning~\cite{Trummer2021}. The authors propose a simple approach based on supervised learning. The approach is trained via tuning hints that have been manually labeled with hint translations. In contrast to that, DB-BERT uses unlabeled text as input. No manual pre-processing is required on this input text. Choices associated with hint translation steps are annotated with manually provided text labels (15~labels in total). However, those labels are not scenario-dependent and we use the same labels across all experiments (Table~\ref{tab:labels} shows five out of the 15 labels). The same applies to all other tuning parameters introduced in the following sections. Besides the differences in manual labeling overheads, the prior approach is purely based on input text, does not integrate any performance measurements, and is therefore unable to adapt recommendations to specific benchmarks or metrics. We discuss differences to prior work in Section~\ref{sec:related} in more detail. 

%We are only aware of one recent vision paper, aimed at leveraging text documents for database tuning~\cite{Trummer2021}. The authors propose a simple approach based on supervised learning, trained via annotated natural language tuning hints. \rev{Hence, the approach requires manually labeling training samples. Annotations must be generated manually. The approach exploits annotated natural language tuning hints as training samples which must } The approach is purely based on input text, does not integrate any performance measurements, and is therefore unable to adapt recommendations to specific benchmarks or metrics. We discuss differences to our work in Section~\ref{sec:related} in more detail. 

In our experiments, we compare against the latter work as well as against state of the art methods for database tuning without input text. We exploit large document collections, mined by issuing Google queries with relevant keywords, as text input for DB-BERT. We consider different benchmarks (TPC-C and TPC-H), metrics (throughput and latency), and database systems (MySQL and Postgres). The experiments demonstrate clearly that DB-BERT benefits significantly from information gained via text analysis.  In summary, our original, scientific contributions are the following:

% The experiments show that information, gained by text analysis, is helpful to speed up the database tuning process. 

% Also, we compare DB-BERT to that prior work in various experiments (see Section~\ref{sec:experiments}), showing that DB-BERT finds better solutions with less tuning time.

% In our experiments, we also compare against state of the art methods for DBMS parameter tuning that do not use any text documents as input. We consider different benchmarks (TPC-C and TPC-H), metrics (throughput and latency), and database systems (MySQL and Postgres). We analyze the performance impact of several of DB-BERT's features, and vary data and document sizes. The experiments show that information, gained by text analysis, is helpful to speed up the database tuning process.

\begin{itemize}
    \item We introduce DB-BERT, a system that combines natural language text documents and run time feedback of benchmark evaluations to guide database tuning.
    \item We describe the mechanisms used by DB-BERT to extract, prioritize, translate, aggregate, and evaluate tuning hints.
    \item We evaluate DB-BERT experimentally and compare against baselines, using multiple benchmarks, metrics, and database systems.
\end{itemize}

The reminder of this paper is organized as follows. We cover required background in learning and NLP in Section~\ref{sec:related}. Then, in Section~\ref{sec:model}, we introduce our problem model and terminology. We give an overview of DB-BERT in Section~\ref{sec:overview}. Then, in Section~\ref{sec:extract}, we describe how DB-BERT extracts and prioritizes candidate hints from text documents. We show how DB-BERT translates single hints in Section~\ref{sec:translate} and how it aggregates and evaluates hints in Section~\ref{sec:aggregate}. In Section~\ref{sec:experiments}, we report experimental results before we conclude with Section~\ref{sec:conclusion}.

%This approach extracts a single set of tuning recommendations from a collection of text documents. It does not integrate any run time feedback and does not adapt recommendations to specific workloads or metrics. Also, it does not feature any mechanisms to resolve conflicting tuning hints across sources and cannot deviate from the precise tuning hints proposed. We present DB-BERT (
\section{Background and Related Work}
\label{sec:related}

We discuss technologies that DB-BERT is based upon. Also, we describe prior work addressing similar problems as DB-BERT.

\subsection{Pre-Trained Language Models}

The field of NLP has recently seen significant progress across a range of long-standing problems~\cite{Wolf2020a}. This progress has been enabled, in particular, by the emergence of large, pre-trained language models~\cite{Howard2018}, based on the Transformer architecture~\cite{Vaswani2017}. Such models address two pain points of prior NLP approaches: lack of task-specific training data and bottlenecks in computational resources for training. Language models are trained, using significant computational resources, on tasks for which training data is readily available in large quantities. For instance, masked language modeling~\cite{Devlin2019a} (i.e., predicting masked words in a sentence) can use arbitrary Web text for training. Instead of training new models from scratch for other NLP-related tasks, pre-trained models can be used as a starting point. This approach tends to reduce the amount of training samples needed, as well as computational training overheads, by orders of magnitude~\cite{Howard2018}. The Transformer architecture~\cite{Vaswani2017} has contributed to this development by enabling massively parallel training of large models with hundreds of millions~\cite{Devlin2019} to hundreds of billions~\cite{Floridi2020} of parameters. DB-BERT, true to its name, uses BERT~\cite{Devlin2019}, one of the most widely used Transformer models at this point.

\subsection{Natural Language Query Interfaces}
\label{sub:nlqi}

Natural language query interfaces~\cite{Herzig2019, Karagiannis2020a, Lin2020} are the most popular application of pre-trained models in the context of databases. At the time of writing, corresponding approaches constitute the state of the art for text-to-SQL translation benchmarks such as WikiSQL~\cite{Zhong2017} or SPIDER~\cite{Yu2020c}. The problem of translating text into queries shares certain characteristics with the problem of extracting tuning hints from text. In both cases, text is translated into a formal representation. However, whereas text-to-SQL methods typically translate a single sentence into one single SQL query, DB-BERT extracts multiple tuning hints from multi-sentence text passages. Also, DB-BERT must aggregate and prioritize conflicting hints obtained from multiple sources (a sub-problem that does not appear in the context of natural language query interfaces). Unlike most prior work on text-to-SQL translation, DB-BERT does not assume the presence of labeled training samples. 

%In the context of database systems, pre-trained language models have been primarily used for building natural language query interfaces~\cite{Herzig2019, Karagiannis2020a, Lin2020}.

%While pre-trained language models have been used in the context of databases before, they are typically used to make data access more user-friendly (e.g., via natural language query interfaces~\cite{Herzig2019, Karagiannis2020a, Lin2020}). Here, we use them to make data processing more efficient instead.

\subsection{Reinforcement Learning}

Reinforcement learning~\cite{Sutton2018} addresses scenarios such as the following. An agent explores an environment, selecting actions based on observations. Those actions may influence the environment (whose dynamics are initially unknown to the agent) and result in reward values. The goal of the agent is to maximize reward, accumulated over time. In order to do so, the agent needs to balance exploration (trying out action sequences about which little is known) with exploitation (exploiting action sequences that seem to work well, based on observations so far). The area of reinforcement learning has produced various algorithms that balance this tradeoff in a principled manner. Specifically, DB-BERT uses the Double Deep Q-Networks~\cite{VanHasselt2016} algorithm. This algorithm learns to estimate action values in specific states via deep learning, using two separate models for selecting actions and evaluating them.

Reinforcement learning has been used for various problems in the database domain~\cite{Basu2015a, Hilprecht2020, Yang2020, Zhang2019a}, including tuning problems (discussed in detail next). Different from prior work, we combine reinforcement learning with NLP to find promising parameter settings. More broadly, our work connects to prior work on leveraging text for reinforcement learning, in particular prior work on instruction following~\cite{Luketina2019a}. However, prior work does not consider performance tuning, specifically database tuning, as we do.

\subsection{Database Tuning}

\begin{table}[t]
    \centering
    \caption{Comparing DB-BERT to prior work on NLP-enhanced database tuning.}
    \begin{tabular}{lll}
        \toprule[1pt]
        \textbf{Criterion} & \textbf{Prior-Main} & \textbf{This} \\
        \midrule[1pt]
            Learning Type & Supervised & Reinforcement Learning \\
            NLP Type & Classification & Multiple Choice \\
            Input & Text & Text + Evaluations \\
            Implicit References & No & Yes \\
            Adapting Hints & No & Yes \\
            Iterative & No & Yes \\
            \bottomrule[1pt]
    \end{tabular}
    \label{tab:comparison}
\end{table}

A recent vision paper~\cite{Trummer2021} on NLP-enhanced database tuning relates most to our work. The prior work trains a Transformer model to recognize sentences containing tuning hints via supervised learning. For sentences classified as tuning hints, it extracts parameters and values according to a simple heuristic. This approach uses only text input but no run time feedback. It extracts a fixed set of recommendations from a document collection, without being able to adapt to specific workloads and performance metrics. DB-BERT, on the other hand, uses hints extracted from text merely as a starting point. It supports a broader range of tuning hints (e.g., implicit hints) and does not require annotated tuning hints during training. We summarize some of the differences in Table~\ref{tab:comparison} and compare both approaches experimentally in Section~\ref{sec:experiments}.

Machine learning is nowadays the method of choice for various database optimization problems, ranging from query optimization~\cite{Krishnan2018, Marcus2018a, Ortiz2018a, skinnerDB} over physical design decisions~\cite{Yang2020, Ding2019, Hilprecht2020} up to database system parameter tuning~\cite{Li2018, Pavlo2017, Zhang2019a}. We address an extended version of the latter problem, expanding the input by natural language text documents.
\section{Problem Model}
\label{sec:model}

We tune configurations for database system parameters.
% (our current implementation considers integer, Boolean, and numeric tuning parameters)
\begin{definition}[Configuration]
Each DBMS is associated with a set $\mathcal{P}$ of configuration parameters. Denote by $\mathcal{V}$ the set of admissible parameter values. A configuration assigns each parameter to a valid value and is represented as a function $\mathcal{P}\mapsto\mathcal{V}$. Equivalently, we represent this function as set $\{\langle p_i,v_i\rangle\}$ for $p_i\in\mathcal{P}$ and $v_i\in\mathcal{V}$ of parameter-value pairs. Parameters not referenced in a configuration maintain their default values.
\end{definition}

Our goal is to find configurations that optimize performance. Traditionally, the following problem model is used.

\begin{definition}[Database Tuning]
A database tuning problem is described by a tuple $\langle b,\mathcal{P},\mathcal{V}\rangle$. Here, $b$ is a benchmark defining a set of queries (or a transaction workload), together with a performance metric to optimize (e.g., run time or throughput). A solution assigns parameters $\mathcal{P}$, selected for tuning, to values from $\mathcal{V}$ and ideally optimizes performance according to benchmark $b$.
\end{definition}

In this work, we address a variant of this problem model.

\begin{definition}[NLP-Enhanced Tuning]
An NLP-enhanced database tuning instance is described by a tuple $\langle b,T,S\rangle$. Here, $b$ is a benchmark to optimize and $T$ a collection of text documents containing tuning hints. The goal is to find optimal configurations for $b$, considering all DBMS tuning knobs (more precisely, our current implementation considers all integer, numeric, and Boolean parameters for each system), using tuning hints extracted from $T$ via natural language analysis. $S$ is a vector of numerical system properties (such as the amount of RAM or the number of cores) needed to translate hints, potentially containing relative value suggestions, into concrete values.
\end{definition}

We do not expect users to specify parameters to tune nor to suggest value ranges for parameters. We rely on natural language analysis to identify relevant parameters and proposed values. However, the approach presented in this work assumes access to a DBMS instance. Via this interface, we verify whether extracted parameter names are valid and whether the parameter type falls within our scope (our current implementation considers integer, Boolean, and numeric parameters). %The goal of text analysis is to extract tuning hints, described next.

\begin{definition}[Tuning Hint]\label{def:hint}
A tuning hint suggests a value for one DBMS parameter. We model tuning hints as a triple $\langle t,p,v\rangle$ where $t$ is a text snippet containing the hint, $p$ a specific parameter, and $v$ a specific value mentioned in $t$. We call the hint explicit if $p$ is mentioned explicitly in $t$ and implicit otherwise. In pseudo-code, we use notation $h.p$ or $h.t$ to refer to parameter or text of hint $h$.
\end{definition}

Note that a text snippet $t$ may contain suggestions for multiple parameters or multiple suggested values for the same parameter. This is why we need $p$ and $v$ to identify a specific hint within $t$. Value $v$ may not always be the concrete value proposed for $p$. This is why we translate tuning hints into formulas, defined next.

\begin{definition}[Translated Hint]\label{def:translated}
We translate tuning hints $\langle t,p,v\rangle$ into a formula of the form $p=f(v,S)$ where $f$ is a formula and $S$ a vector of numerical system properties (e.g., the amount of main memory). We consider formulas of type $f(v,S)=v\cdot m$ as well as $f(v,S)=v\cdot S_i\cdot m$ where $S_i$ is the $i$-th component of $S$ and $m\in \mathbb{R}$ a multiplicator (picked from a discrete set $M$ of multiplicators). 
\end{definition}

We illustrate tuning hints and their translation.

\begin{example}
Consider the text snippet $t=$``Properly configure \verb|shared_buffers| - we recommend 25\% of available RAM''\footnote{\url{https://blog.timescale.com/blog/13-tips-to-improve-postgresql-insert-performance/}}. Assume $S=\langle 8GB, 4, 1TB\rangle$ describes the amount of RAM, the number of cores, and the amount of disk space on the target system. Then, the tuning hint $\langle t,p,v\rangle$ for $p=$\verb|shared_buffers| and $v=0.25$ should translate into the formula $f(v,S)=v\cdot S_0\cdot 1$ (where 1 represents the multiplicator), which evaluates to 2~GB.
\end{example}
\section{System Overview}
\label{sec:overview}

% \tikzstyle{system}=[fill=red!10, rectangle, draw]
% \tikzstyle{header}=[font=\bfseries]
% \tikzstyle{component}=[fill=blue!20, align=center, rectangle, draw, minimum width=1.5cm]
% \tikzstyle{io}=[font=\itshape]
% \tikzstyle{dataflow}=[thick, ->]

% \begin{figure}
%     \centering
%     \begin{tikzpicture}
%         \node[system, minimum width=6cm, minimum height=1.5cm] (dbbert) at (2,0.2) {};
%         \node[header] at (2,0.7) {DB-Bert};
%         \node[io, yshift=0.5cm] (input) at (dbbert.north) {Benchmark, Hardware Properties, Text documents};
%         \node[io, yshift=-0.5cm] (output) at (dbbert.south) {Recommended configuration};
%         \node[component] (extract) at (0,0) {Extract\\Hints};
%         \node[component] (translate) at (2,0) {Create\\Config.};
%         \node[component] (evaluate) at (4,0) {Evaluate\\Config.};
%         \node[component] (dbms) at (6,0.2) {DBMS};
%         \draw[dataflow] (extract) -- (translate);
%         \draw[dataflow] (translate) -- (evaluate);
%         \draw[dataflow, <->] (dbbert) -- (dbms);
%         \draw[dataflow] (input) -- (dbbert);
%         \draw[dataflow] (dbbert) -- (output);
%         \draw[dataflow] (evaluate.north) to[out=170, in=10] (translate.north);
%     \end{tikzpicture}
%     \caption{Overview of DB-BERT system: we exploit tuning hints, extracted from text documents, to find optimal DBMS knob settings for a given workload.}
%     \label{fig:overview}
% \end{figure}

\tikzstyle{system}=[fill=red!10, rectangle, draw]
\tikzstyle{header}=[font=\bfseries]
\tikzstyle{component}=[fill=blue!20, align=center, rectangle, draw, minimum width=3.5cm, rounded corners=1pt]
\tikzstyle{io}=[font=\itshape]
\tikzstyle{dataflow}=[thick, ->]
\tikzstyle{compname}=[font=\bfseries\itshape]
\newcommand\comp[1]{{\textbf{#1}}}

\tikzstyle{system}=[draw, rectangle, rounded corners=1mm, shade, top color=gray!10, bottom color=gray!20, blur shadow={shadow blur steps=5}]
\tikzstyle{component}=[draw, rectangle, shade, top color=blue!10, bottom color=blue!20, minimum width=3.7cm, font=\bfseries, blur shadow={shadow blur steps=5}, align=center]

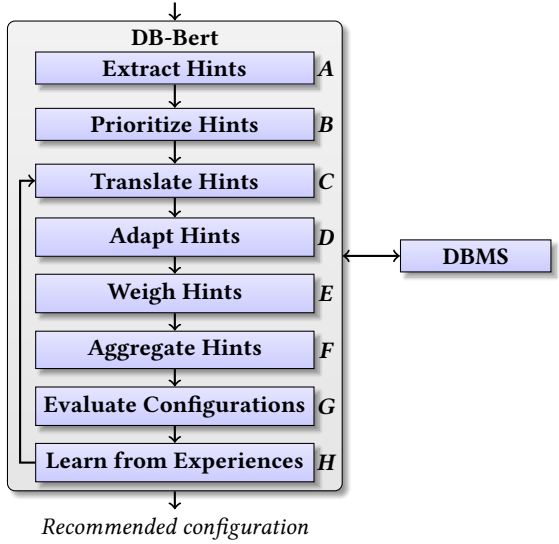
\begin{figure}
    \centering
    \begin{tikzpicture}
        \node[system, minimum width=4.45cm, minimum height=6.25cm] (dbbert) at (2,-0.5) {};
        \node[header] at (2,2.4) {DB-Bert};
        \node[io, yshift=0.5cm] (input) at (dbbert.north) {Benchmark, Hardware Properties, Text documents};
        \node[io, yshift=-0.5cm] (output) at (dbbert.south) {Recommended configuration};
        \node[component] (extract) at (2,2) {Extract Hints};
        \node[component] (prioritize) at (2,1.25) {Prioritize Hints};
        \node[component] (translate) at (2,0.5) {Translate Hints};
        \node[component] (adapt) at (2,-0.25) {Adapt Hints};
        \node[component] (weigh) at (2,-1) {Weigh Hints};
        \node[component] (aggregate) at (2,-1.75) {Aggregate Hints};
        \node[component] (evaluate) at (2,-2.5) {Evaluate Configurations};
        \node[component] (learn) at (2,-3.25) {Learn from Experiences};
        \node[component, minimum width=2cm] (dbms) at (6,-0.5) {DBMS};
        
        \node[compname] at (4,2) {A};
        \node[compname] at (4,1.25) {B};
        \node[compname] at (4,0.5) {C};
        \node[compname] at (4,-0.25) {D};
        \node[compname] at (4,-1) {E};
        \node[compname] at (4,-1.75) {F};
        \node[compname] at (4,-2.5) {G};
        \node[compname] at (4,-3.25) {H};
        
        \draw[dataflow] (extract) -- (prioritize);
        \draw[dataflow] (prioritize) -- (translate);
        \draw[dataflow] (translate) -- (adapt);
        \draw[dataflow] (adapt) -- (weigh);
        \draw[dataflow] (weigh) -- (aggregate);
        \draw[dataflow] (aggregate) -- (evaluate);
        \draw[dataflow] (evaluate) -- (learn);
        \draw[dataflow] (learn.west) -- ++ (-0.2,0) -- ++ (0,3.75) -- (translate.west);
        \draw[dataflow, <->] (dbbert) -- (dbms);
        \draw[dataflow] (input) -- (dbbert);
        \draw[dataflow] (dbbert) -- (output);
        % \draw[dataflow] (evaluate.north) to[out=170, in=10] (translate.north);
    \end{tikzpicture}
    \caption{Overview of DB-BERT system: we exploit tuning hints, extracted from text documents, to find optimal DBMS knob settings for a given workload.}
    \label{fig:overview}
\end{figure}

Figure~\ref{fig:overview} shows an overview of the DB-BERT system. DB-BERT searches settings for the tuning knobs of a DBMS that maximize performance according to a specific benchmark (specifying workload and performance metric). DB-BERT differs from prior tuning system in that it exploits text documents about the DBMS to tune, for instance the DBMS manual, as additional input.

DB-BERT obtains as input the benchmark to tune, a collection of text documents containing suggested settings for tuning knobs, and numerical properties describing the hardware platform (namely, our implementation expects the amount of RAM, the number of cores, and the amount of disk space as inputs). The latter input is necessary to translate tuning hints in text documents that use \textit{relative} recommendations (e.g., suggesting a buffer size as a percentage of the amount of RAM). Note that DB-BERT is not restricted to parameters that relate to the aforementioned hardware properties. DB-BERT can process hints for arbitrary parameters, as long as recommended values are specified as \textit{absolute} values in text.

DB-BERT does not use text input alone to determine parameter settings (separating it from prior work on NLP-enhanced database tuning~\cite{Trummer2021}). Instead, it exploits run time feedback obtained by benchmarking specific configurations on the DBMS to tune. Hence, DB-BERT requires a connection to a DBMS instance.

At the start of a tuning session, DB-BERT divides input text into text snippets and tries to extract tuning hints from each snippet (Step \comp{A} in Figure~\ref{fig:overview}). A tuning hint corresponds to a recommendation of a specific value for a specific parameter. Extracting hints from text snippets is non-trivial, in particular as parameter references may be implicit (i.e., the text does not explicitly mention the name of the parameter to tune). Next, DB-BERT determines the order in which hints will be considered in the following stages (Step \comp{B} in Figure~\ref{fig:overview}). Ideally, the most important hints are considered first. DB-BERT uses a heuristic to order hints, prioritizing hints about frequently mentioned parameters while limiting the number of hints considered consecutively for the same parameter.

Next, DB-BERT iteratively constructs configurations (i.e., value assignments for tuning knobs) from tuning hints. It evaluates those configurations on the input benchmark via trial runs. Iterations continue until the user interrupts optimization or a user-specified optimization time limit is reached.

In each iteration, DB-BERT considers a batch of tuning hints (not the entire set of tuning hints). It considers hints in the order established at the start of the tuning session, thereby considering the seemingly most important hints first. For each hint, DB-BERT takes three types of decisions. First, it translates the hint text into a simple equation, assigning a value to a parameter (Step~\comp{C} in Figure~\ref{fig:overview}). Second, in Step~\comp{D}, it decides whether to deviate from the recommended value (i.e., whether to multiply the recommended value by a constant). Third, it assigns a weight to the hint (Step~\comp{E}). These weights decide how to prioritize in case of conflicting recommendations about the same tuning knob. After treating all hints in the current batch, DB-BERT aggregates them into a small set of configurations (Step~\comp{F}), mediating between inconsistent recommendations using hint weights. It evaluates those configurations on the user-specified benchmark via trial runs (Step~\comp{G} in Figure~\ref{fig:overview}).

DB-BERT learns to improve the way hints are translated, adapted, and weighted over the course of a tuning session. This allows DB-BERT to specialize a configuration to the current benchmark and platform. DB-BERT uses reinforcement learning to make all decisions associated with Steps~\comp{C} to \comp{E} in Figure~\ref{fig:overview}. The learning process is therefore driven by a reward function that the system tries to maximize. In case of DB-BERT, that reward function is based on the performance results for specific configurations during trial runs. Configurations that are accepted by the DBMS (i.e., trying to set parameters to specific values does not result in an error) and achieve high performance generate high reward values. Based on rewards received, the system learns to improve its decision making in coming iterations (Step~\comp{H} in Figure~\ref{fig:overview}).

DB-BERT uses \textit{deep} reinforcement learning. This means that immediate and future reward values associated with specific choices are estimated using a neural network. Specifically, DB-BERT uses BERT, a pre-trained language model, as neural network. Due to pre-training, this model comes with powerful natural language analysis capabilities out of the box. To estimate the value of specific choices during Steps~\comp{C} to \comp{E}, BERT is applied to pairs of text snippets. The first snippet is taken from the text of a tuning hint, the second snippet is a text label representing the semantics of that choice (see Table~\ref{tab:labels} in Section~\ref{sec:translate} for example labels). Based on reward values received, the initial weights of the BERT model are refined over the course of a tuning session (in Step~\comp{H}).

%The tuning process is iterative. In each iteration, DB-BERT selects a configuration to benchmark. To select configurations, DB-BERT uses reinforcement learning. Different from prior work exploiting reinforcement learning for database tuning, DB-BERT does not learn promising configurations from scratch. Instead, it rather learns a \textit{function that interprets tuning documents}. Interpreting tuning hints entails translating natural language tuning hints into concrete tuning suggestions, weighting hints from different sources against each other, and adapting the original hints as needed (e.g., by deviating from proposed values). Using tuning documents, we learn relevant parameters and reasonable value ranges for them. 

\begin{algorithm}[t]
\caption{NLP-enhanced database performance tuning.}
\label{alg:main}
\renewcommand{\algorithmiccomment}[1]{// #1}
\begin{algorithmic}[1]
\State \Comment{Optimize all DBMS parameters $P$ for benchmark $b$ via hints}
\State \Comment{from text collection $T$, using multiplicators $M$, system}
\State \Comment{properties $S$, and weights $W$ for translating hints. Use up}
\State \Comment{to $l$ hints per parameter and episode and up to $e$ hints}
\State \Comment{per episode. Evaluate $n$ configurations in each episode.}
\Function{DB-BERT}{$T,b,P,M,S,W,l,e,n$}
\State \Comment{Extract tuning hints from text snippets}
\State $H\gets\cup_{t\in T}$\Call{ExtractHints}{$P,t$}
\State \Comment{Order tuning hints by priority}
\State $H_o\gets$\Call{OrderHints}{$H,l$}
\While{No timeout}
\State \Comment{Iterate over hints in priority order}
\For{$H_b\gets$\textproc{Batches}($H_o,e$)}
\State \Comment{Evaluate configurations created using hints}
\State \Call{RunEpisode}{$b,H,S,M,W,n$}
\EndFor
\EndWhile
\State \Return{Best configuration found}
\EndFunction
\end{algorithmic}
\end{algorithm}

Algorithm~\ref{alg:main} represents the main function, executed by DB-BERT, in pseudo-code. The input integrates user-provided inputs, represented in Figure~\ref{fig:overview}, as well as other parameters, extracted automatically or kept constant across systems and benchmarks. These include the full set of integer, Boolean, and numeric tuning knobs, extracted from the DBMS, $P$, a set $M$ of multiplicators (to deviate from values proposed in text), a set $W$ of weights (to determine relative importance between conflicting hints from different sources), and parameters $l$, $e$, and $n$ to choose the number of hints processed per parameter and iteration, the total number of hints considered per iteration, and the number of configurations evaluated per iteration, respectively. The semantics of those parameters will be described in more detail in the following sections.

Line~8 in Algorithm~\ref{alg:main} realizes Step~\comp{A} from Figure~\ref{fig:overview}, Line~10 realizes Step~\comp{B}. 
The main loop iterates until the tuning time budget is depleted. Function~\Call{Batches}{$H_o,e$} divides hints into batches of size at most $e$, following the previously established hint order. Each invocation of \textproc{RunEpisode} realizes Steps~\comp{C} to \comp{H} from Figure~\ref{fig:overview}. Finally, DB-BERT recommends the best observed configuration.

%DB-BERT first extracts candidate tuning hints from the input text. Next, DB-BERT orders hints in order to prioritize hints about frequently mentioned parameters (while ensuring diversity of hints by limiting the number of hints per parameter as well). Extraction and hint ordering are described in Section~\ref{sec:extract}. The main loop iterates until the tuning time budget is depleted. In each iteration, DB-BERT considers tuning hints in batches, following the previously established order. Function~\Call{Batches}{$H_o,e$} divides hints into batches of size at most $e$, following the previously established hint order. 

%For each batch of hints, DB-BERT performs one reinforcement learning episode. Each episode starts from an empty configuration (i.e., using default values for all tuning knobs). Then, the empty configuration is updated using current hints. This process entails applying BERT, a pre-trained language model, to each natural language hint. The goal is to translate each hint into a parameter and value formula (which may depend on values provided in text as well as on hardware properties), possibly adapting the suggested values, and assigning hints to weights. The reward function rewards performance improvements, compared to the default configuration (and penalizes value assignments rejected by the DBMS). Based on rewards observed, the weights of the BERT model are updated during the tuning process. The learning process is described in more detail in Sections~\ref{sec:translate} and \ref{sec:aggregate}. Finally, DB-BERT recommends the best observed configuration.

Section~\ref{sec:extract} discusses hint extraction and ordering. Section~\ref{sec:translate} describes the learning process in more detail and Section~\ref{sec:aggregate} outlines how hints are aggregated into configurations.
\section{Extracting Candidate Hints}
\label{sec:extract}

\begin{algorithm}[t]
\caption{Extract candidate tuning hints from text documents.}
\label{alg:extraction}
\renewcommand{\algorithmiccomment}[1]{// #1}
\begin{algorithmic}[1]
\State \Comment{Extract tuning hints about parameters $P$ from text $t$.}
\Function{ExtractHints}{$P,t$}
\State \Comment{Extract explicit parameter references}
\State $E\gets\{p\in P|contains(t,p)\}$
\State \Comment{Extract implicit parameter references}
\State $i\gets\arg\min_{p\in P}\delta(BERT(p),BERT(t))$
\State \Comment{Extract candidate parameter values}
\State $V\gets$\Call{ExtractValues}{$t$}$\cup\{0,1\}$
\State \Comment{Return pairs of values and parameters}
\State \Return{$\{\langle t,p,v\rangle|p\in E\cup\{i\},v\in V\}$}
\EndFunction
\end{algorithmic}
\end{algorithm}

\tikzstyle{estep}=[component, minimum width=1.6cm, minimum height=0.8cm]

\begin{figure}
    \centering
    \begin{tikzpicture}
        \node[io] (parameters) at (0,1.2) {Parameters};
        \node[io] (passage) at (4,1.2) {Text passage};
        \node[estep] (pbert) at (0,0) {Encode\\via BERT};
        \node[estep] (tbert) at (2,0) {Encode\\via BERT};
        \node[estep] (extractp) at (4,0) {Extract\\Parameters};
        \node[estep] (extractv) at (6,0) {Extract\\Values};
        \node[estep] (similar) at (0,-1.2) {Cosine\\Similarity};
        \node[estep] (topk) at (2,-1.2) {Top-K\\Matches};
        \node[estep] (union) at (4,-1.2) {$\cup$};
        \node[estep] (cross) at (6,-1.2) {$\times$};
        \node[io] (relationships) at (3,-2.4) {Candidate hints};
        \draw[dataflow] (parameters) -- (pbert.north);
        \draw[dataflow] (passage) -- (tbert.north);
        \draw[dataflow] (passage) -- (extractp.north);
        \draw[dataflow] (passage) -- (extractv.north);
        \draw[dataflow] (pbert) -- (similar);
        \draw[dataflow] (tbert) -- (similar);
        \draw[dataflow] (similar) -- (topk);
        \draw[dataflow] (topk) -- (union);
        \draw[dataflow] (extractp) -- (union);
        \draw[dataflow] (extractv) -- (cross);
        \draw[dataflow] (union) -- (cross);
        \draw[dataflow] (cross) -- (relationships);
        \draw[dataflow] (parameters) -- (extractp.north);
    \end{tikzpicture}
    \caption{Given a text passage and DBMS parameter names, DB-BERT pairs extracted values with parameters that are explicitly mentioned or are similar to the text.}
    \label{fig:extraction}
\end{figure}
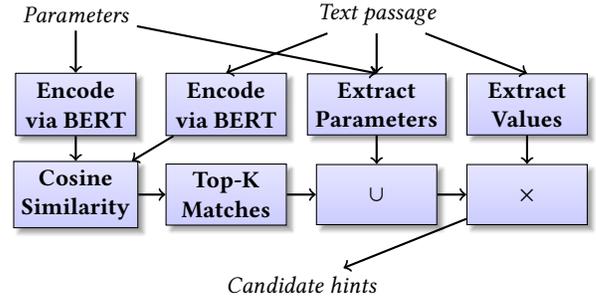

In a first step, DB-BERT extracts candidate tuning hints. Following Definition~\ref{def:hint}, a tuning hint consists of a text snippet, a parameter reference, and a value reference. Algorithm~\ref{alg:extraction} describes the extraction process (illustrated in Figure~\ref{fig:extraction} as well). It extracts explicit as well as \textit{implicit} parameter references. Implicit references are obtained by comparing the BERT encoding for the text (a vector) against BERT encodings of parameter names, selecting the parameter with minimal cosine distance. We consider all numbers that appear in text, potentially combined with size units, as potential value suggestions. By default, we add values 0 and 1, representing on and off values for Boolean flags, into the set of values (on and off values are often not explicitly mentioned in tuning hints). The set of candidate hints for a given text snippet is the Cartesian product between parameter references and values. This means that our candidates likely contain \textit{erroneous} hints (i.e., parameter-value combinations that are not linked by the text). The task of separating actual from erroneous hints is solved during the translation phase, described in the next section.

\begin{algorithm}[t]
\caption{Prioritize hints based on their parameters.}
\label{alg:striding}
\renewcommand{\algorithmiccomment}[1]{// #1}
\begin{algorithmic}[1]
\State \Comment{Order hints $H$ using stride of length $l$.}
\Function{OrderHints}{$H,l$}
\State \Comment{Collect parameters in hints}
\State $P\gets\{h.p|h\in H\}$
\State \Comment{Group hints by parameter}
\State $G\gets\{\langle p_i,H_i\rangle|H=\dot{\cup} H_i,\forall i\forall h\in H_i:h.p=p_i,P=\dot{\cup} P_i\}$
\State \Comment{Sort parameters by hint count}
\State $p_0,\ldots,p_n\gets P$ sorted by number of hints (ascending)
\State \Comment{Initialize result list}
\State $R\gets[]$
\State \Comment{Iterate over hint ranges}
\For{$i\gets 0,\ldots,\lceil |G(p_0)/l|\rceil$}
\State \Comment{Iterate over (ordered) parameters}
\For{$p\gets p_0,\ldots,p_n$}
\State \Comment{Add hints on $p$ within $i$-th range}
\State \Call{Append}{$R,G(p)[i\cdot l:(i+1)\cdot l-1]$}
\EndFor
\EndFor
\State \Return{$R$}
\EndFunction
\end{algorithmic}
\end{algorithm}

%\tikzstyle{hints}=[fill=blue!10, align=left, draw, anchor=west]
\tikzstyle{hints}=[component, align=left, draw, anchor=west]
\tikzstyle{considerhint}=[red, ultra thick, ->, draw]
\tikzstyle{jumptohint}=[red, dashed, ->, draw]
\tikzstyle{stridenr}=[circle, fill=red, text=white]

\begin{figure}
    \centering
    \begin{tikzpicture}
        \node[hints, minimum width=8cm, text width=7.8cm] at (0,0) {Parameter 1 Hints};
        \node[hints, minimum width=6cm, text width=5.8cm] at (0,-1) {Parameter 2 Hints};
        \node[hints, minimum width=3cm, text width=2.8cm] at (0,-2) {Parameter 3 Hints};
        \draw[considerhint] (0,0.5) -- (4,0.5);
        \draw[considerhint] (0,-0.5) -- (4,-0.5);
        \draw[considerhint] (0,-1.5) -- (3,-1.5);
        \draw[considerhint] (4.1,0.5) -- (8,0.5);
        \draw[considerhint] (4.1,-0.5) -- (6,-0.5);
        \node[stridenr] at (0,0.5) {1};
        \node[stridenr] at (0,-0.5) {2};
        \node[stridenr] at (0,-1.5) {3};
        \node[stridenr] at (4.3,0.5) {4};
        \node[stridenr] at (4.3,-0.5) {5};
    \end{tikzpicture}
    \caption{DB-BERT prioritizes hints about frequently mentioned parameters while limiting the number of hints per parameters before switching to the next one. In the illustrated example, hints are considered in the order indicated by the red (numbered) arrows.}
    \label{fig:hintorder}
\end{figure}
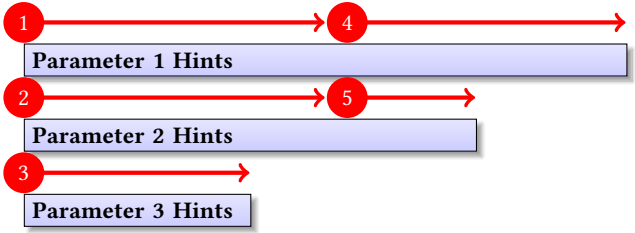

After extracting candidate hints, DB-BERT sorts them using Algorithm~\ref{alg:striding}. Our goal is to increase chances of finding promising configurations when considering hints in sort order. We consider two rules of thumb. First, we expect important parameters to be mentioned in more documents. Second, we expect diminishing returns when considering more and more hints about the same parameter. As a result, we prioritize hints about parameters that appear in more documents. However, we consider at most a fixed number of hints about the same parameter, before switching to the next one. Algorithm~\ref{alg:striding} implements those high-level principles. After grouping hints by parameter, it iterates over hint index ranges. For each index range, it iterates over parameters in decreasing order of occurrences, adding up to $l$ hints per parameter before switching to the next one (until no new hints are left to add for any parameter). %The following example illustrates the sort order.

\begin{example}
Figure~\ref{fig:hintorder} illustrates hint ordering with three parameters. Blue rectangles represent hints for each parameter. The horizontal width is proportional to the number of hints. Starting with the most frequently mentioned parameter, we add a limited number of hints for each parameter. After treating the least frequently mentioned parameter (symbolized by the red arrow),  Parameter~3, we start again with the first one until no more hints are left.
\end{example}
\section{Translating Single Hints}
\label{sec:translate}

\begin{comment}
\begin{algorithm}
\caption{Transition function for translating single hints.}
\label{alg:singlestep}
\renewcommand{\algorithmiccomment}[1]{// #1}
\begin{algorithmic}[1]
\State \Comment{For benchmark $b$, translate hint linking parameter $p$ to}
\State \Comment{value $v$, using system properties $S$ and multiplicators $M$.}
\State \Comment{Action $a$ refers to decision $d$ and expands partial formula $f$.}
\State \Comment{Returns next decision, expanded formula, and reward value.}
\Function{TStep}{$b,p,v,S,M,d,f,a$}
\If{$d=0$}
\State \Comment{Decide hint type and whether to use it}
\If{$a=\mathrm{NO\_HINT}$}
\State \Return{$\langle\mathrm{END},-,0\rangle$}
\Else
%\State $f\gets v\cdot P_a$
\State \Return{$\langle d+1,v\cdot S_a,0\rangle$}
\EndIf
\ElsIf{$d=1$}
\State \Comment{Choose multiplicator for current hint value}
%\State $f\gets f\cdot M_a$
\State \Return{$\langle d+1,f\cdot M_a,0\rangle$}
\Else
\State \Comment{Try setting parameter value and benchmark on $b$}
\State $suc\gets$\Call{DbmsSet}{$p,f$}
\If{$suc=\mathrm{True}$}
\State $r\gets$\Call{EvaluatePerformance}{$b$}
\State \Return{$\langle\mathrm{END},f,r+1\rangle$}
\Else
\State \Return{$\langle\mathrm{END},-,-1\rangle$}
\EndIf
\EndIf
\EndFunction
\end{algorithmic}
\end{algorithm}
\end{comment}

\begin{algorithm}[t]
\caption{Transition function for translating single hints.}
\label{alg:singlestep}
\renewcommand{\algorithmiccomment}[1]{// #1}
\begin{algorithmic}[1]
\State \Comment{For benchmark $b$, translate hint linking parameter $p$ to}
\State \Comment{value $v$, using system properties $S$ and multiplicators $M$.}
\State \Comment{Action $a$ refers to decision $d$ and expands partial formula $f$.}
\State \Comment{Returns next decision, expanded formula, and reward value.}
\Function{TStep}{$b,p,v,S,M,d,f,a$}
\If{$d=0$}
\State \Comment{Decide hint type and whether to use it}
\If{$a=\mathrm{NO\_HINT}$}
\State \Return{$\langle\mathrm{END},-,0\rangle$}
\Else
%\State $f\gets v\cdot P_a$
\State \Return{$\langle d+1,v\cdot S_a,0\rangle$}
\EndIf
\ElsIf{$d=1$}
\State \Comment{Choose multiplicator for current hint value}
\State $f\gets f\cdot M_a$
\State \Comment{Try setting parameter value and benchmark on $b$}
\State $suc\gets$\Call{DbmsSet}{$p,f$}
\If{$suc=\mathrm{True}$}
\State $r\gets$\Call{EvaluatePerformance}{$b$}
\State \Return{$\langle\mathrm{END},f,r+1\rangle$}
\Else
\State \Return{$\langle\mathrm{END},-,-1\rangle$}
\EndIf
\EndIf
\EndFunction
\end{algorithmic}
\end{algorithm}

DB-BERT translates tuning hints into arithmetic formulas (see Definition~\ref{def:translated} for details). Those formulas may depend on values, specified in text, as well as on system properties such as the amount of main memory. Evaluating a formula yields a value suggestion for a tuning knob. 

For each tuning hint, we model the translation as a sequence of decisions. We learn to translate tuning hints by using reinforcement learning. Reinforcement learning is generally applied to Markov Decision Processes (MDPs), specified by a set of states, actions, a transition function mapping state and action pairs to new states, and a reward function. A reinforcement learning agent learns to make decisions maximizing expected rewards, using observations as guidance. In our scenario, states represent (partially specified) arithmetic formulas. Actions specify parts of the formula. The transition functions links partially specified formulas and actions to states representing the formula, completed as specified in the action. The reward function is based on feedback from the DBMS, penalizing translations that result in inadmissible configurations while rewarding changes that improve performance. We describe the structure of the environment (i.e., states, actions, transitions, and rewards) in Section~\ref{sub:environment} and the structure of the learning agent in Section~\ref{sub:agent}.

\subsection{Learning Environment}
\label{sub:environment}

%\tikzstyle{state}=[fill=red!10, minimum width=1.5cm, minimum height=0.7cm, draw]
\tikzstyle{state}=[draw, rectangle, shade, top color=blue!10, bottom color=blue!20, blur shadow={shadow blur steps=5}, minimum width=1.5cm, minimum height=0.7cm, draw]
\tikzstyle{endstate}=[state, double]
\tikzstyle{transition}=[dataflow]

\begin{figure}
    \centering
    \begin{tikzpicture}
        \node[io] (start) at (0,0) {Text passage; Parameter $p$; Value $v$};
        \node[endstate] (nohint) at (-3.25,-1.5) {-};
        \node at (-1.6,-1.5) {\ldots};
        \node[state] (reshint) at (0,-1.5) {$p=v\cdot S_i$};
        \node at (1.6,-1.5) {\ldots};
        \node[state] (rawhint) at (3.25,-1.5) {$p=v$};
        \node at (-3,-3) {\ldots};
        \node[state] (modhint) at (0,-3) {$p=v\cdot S_i\cdot M_j$};
        \node at (3,-3) {\ldots};
        \node at (-3.5,-4.5) {\ldots};
        \node[endstate] (error) at (-1.5,-4.5) {DBMS: Error};
        \node[state] (valid) at (1.5,-4.5) {DBMS: Valid};
        \node at (3.5,-4.5) {\ldots};
        \node[endstate] (evaluated) at (0,-6) {Evaluated};
        \draw[transition] (start) -- (nohint);
        \draw[transition] (start) -- (reshint);
        \draw[transition] (start) -- (rawhint);
        \draw[transition] (reshint) -- (modhint);
        \draw[transition, dashed] (modhint) -- (error);
        \draw[transition, dashed] (modhint) -- (valid);
        \draw[transition] (valid) -- (evaluated);
        \node at (-3,-0.5) {No Hint};
        \node at (3,-0.5) {Absolute};
        \node at (0.75,-0.75) {Relative};
        \node at (1.25,-2.25) {Modify Value};
        \node[align=center] at (0,-3.75) {Try\\Setting};
        \node at (2.5,-5.25) {Run Benchmark};
    \end{tikzpicture}
    \caption{Markov Decision Process for hint translation: parameter-value pairs are mapped to formulas by action sequences. Rectangles represent states (double lines mark end states). Arrows represent transitions (dashed arrows mark non-deterministic transitions).}
    \label{fig:translation}
\end{figure}
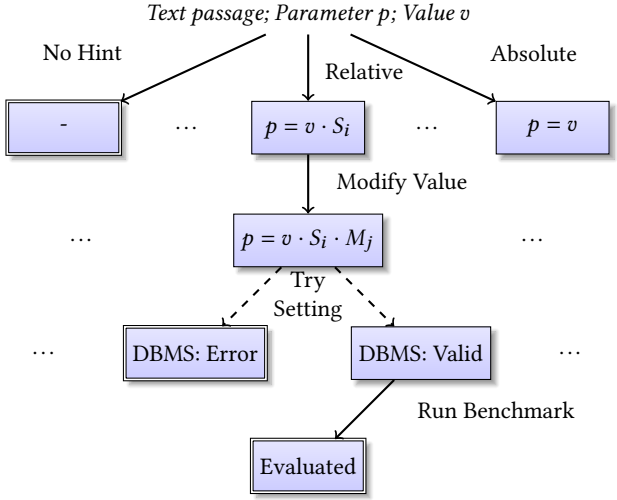
%\end{comment}

Algorithm~\ref{alg:singlestep} implements the transition function, used by DB-BERT to translate single hints (the pseudo-code is close to the implementation of the step function in the corresponding OpenAI Gym environment\footnote{\url{https://gym.openai.com/}}). In Algorithm~\ref{alg:singlestep}, and for a fixed tuning hint, the current state is characterized by a partially specified formula ($f$) and by variable $d$, the integer ID of the next decision to take. For each hint, we start with an empty formula $f$ and $d=0$. We represent actions (input $a$) as integer numbers from one to five. The semantics of actions depend on the value of $d$. For $d=0$, the action decides whether the current hint is erroneous (constant NO\_HINT) and, if not, whether the hint suggests a relative or absolute parameter value. Relative values are expressed as percentage of system properties such as main memory or the number of cores (stored in vector $S$ with $S_a$ representing a specific vector component). For relative values, we set $f$ to the product between value $v$ and the corresponding system property. We unify treatment of relative and absolute values by setting $S_1=1$ (i.e., $a=1$ represents an absolute value). 

For $d=1$, the action picks a multiplicator from $M$ that allows deviating from the proposed value. Unlike prior work merely extracting tuning hints~\cite{Trummer2021}, such multiplicators allow DB-BERT to adapt to specific benchmarks. In the next section, we introduce an additional decision that weighs hints. Here, we have fully specified the formula after two decisions. Next, we try setting parameter $p$ to the formula evaluation result. If the setting is rejected by the DBMS, we directly advance to an end state (constant END). This case yields negative reward (motivating our agent to learn translating hints into admissible formulas). Otherwise, we evaluate performance on the input benchmark $b$. The result is a reward value. Higher rewards are associated with better performance. We calculate reward by comparing performance with a configuration to evaluate to performance with default settings. For OLAP benchmarks (e.g., TPC-H), we use the delta of run times (scaled by a constant). For OLTP benchmarks (e.g., TPC-C), we use the throughput delta. 

We reward configurations that are admissible and increase performance. Those two metrics are immediately relevant for tuning. We use them when applying DB-BERT for tuning a specific system for a specific benchmark. Before applying DB-BERT for specific tuning tasks, we perform a training phase to fine-tune DB-BERT's language models for hint translation in general. To speed up convergence, only during training, we add an additional component to the reward function. This component rewards settings that seem more likely, e.g.\ since they are in the same order of magnitude as the default settings for a parameter. Such heuristics replace manually generated hint translations, used in prior work~\cite{Trummer2021}. Figure~\ref{fig:translation} illustrates the MDP behind the translation process (some of the states in Figure~\ref{fig:translation} are not explicitly represented in Algorithm~\ref{alg:singlestep}).

\subsection{Learning Agent}
\label{sub:agent}

%We consider an action space ($a$ is the actionFor $d=0$, the action decides whether the current candidate hint does indeed represent a tuning hint. Also, it decides whether 

\begin{algorithm}[t]
\caption{Evaluating expected reward of actions.}
\label{alg:rewards}
\renewcommand{\algorithmiccomment}[1]{// #1}
\begin{algorithmic}[1]
\State \Comment{Estimate value of action $a$ for decision $d$,}
\State \Comment{given text $t$, parameter $p$ and value $v$.}
\Function{EvaluateAction}{$t,p,v,d,a$}
\State \Comment{Get text associated with choice}
\State $l\gets\mathrm{CHOICE\_LABEL}[d,a]$
\State \Comment{Instantiate label for current hint}
\State $l\gets$\Call{Instantiate}{$l,p,v$}
\State \Comment{Generate input text for BERT}
\State $i\gets t\circ l$
\State \Comment{Generate input types for BERT}
\State $t\gets 0^{|t|}\circ 1^{|l|}$
\State \Comment{Generate input mask for BERT}
\State $u\gets 1^{|i|}$
\If{MASKED\_MODE}
\State $u\gets$\Call{Mask}{$t,p$}
\EndIf
\State \Return{\textproc{BERT}($\langle i,t,u\rangle$)}
\EndFunction
\end{algorithmic}
\end{algorithm}

DB-BERT introduces a learning agent to choose actions in order to maximize rewards. In each state, the agent selects among a discrete set of options. Each option can be expressed as a natural language statement. We can find out which option is correct by comparing that statement against the tuning hint text. Hence, we model action selection as a ``multiple choice question answering problem''. Pre-trained language models can be used to solve this problem (in our implementation, we use the \verb|BertForMultipleChoice| Transformer model\footnote{\url{https://huggingface.co/transformers/model_doc/bert.html}}). We fine-tune model weights during training, based on rewards received.

Algorithm~\ref{alg:rewards} shows how the agent evaluates specific actions, based on observations. Besides the action to evaluate, the input includes a description of the current tuning hint (tuning text $t$, parameter $p$, and value $v$) as well as the current translation step (decision $d$). We associate each combination of an action and a decision with a \textit{label}. The array containing those labels is represented via constant CHOICE\_LABEL in the pseudo-code. The label is a natural language sentence, representing the semantics of the associated choice. It contains placeholders for the concrete parameter and value in the tuning hint. The \Call{Instantiate}{} function replaces placeholders by concrete values. 

\begin{table}[t]
    \centering
    \begin{tabular}{ll}
        \toprule[1pt]
         \textbf{Action} & \textbf{Label} \\
         \midrule[1pt]
         0 (NO\_HINT) & [p] and [v] are unrelated. \\
         1 & [p] and [v] relate to main memory. \\
         2 & [p] and [v] relate to hard disk. \\
         3 & [p] and [v] relate to core counts. \\
         4 & Set [p] to [v]. \\
         \bottomrule[1pt]
    \end{tabular}
    \caption{Labels associated with actions for decision $d=0$. Placeholders are contained in square brackets.}
    \label{tab:labels}
\end{table}

The BERT model uses three inputs: an input text, a type tag associating input tokens with one of two input types, and a mask indicating tokens to consider. Here, we concatenate hint text and instantiated label to form the input text. Types separate hint text from label. By default, all input text is considered for processing. An exception occurs during our generic training phase (see Section~\ref{sub:environment} for more details). Here, we want to avoid learning the names of specific parameters as they do not generalize across systems. Hence, we mask all occurrences of the current parameter name (Function~\textproc{Mask}). On the other side, if learning system and benchmark specific configurations for a concrete tuning problem, there are no reasons to hide information. Algorithm~\ref{alg:rewards} uses a Boolean flag (MASKED\_MODE) to switch between these two modes. 

Table~\ref{tab:labels} shows labels associated with different actions and the first decision level. At this level, we decide whether a candidate hint represents an actual hint and, if so, whether the value is relative or absolute. Finally, we illustrate the translation by an example.

\begin{example}
Consider the tuning hint $\langle t,p,v\rangle$ with $t=$``Set \verb|shared_buffers| to 25\% of RAM'', $p=$\verb|shared_buffers|, and $v=25\%$. First, the agent decides whether the hint is valid and whether it recommends an absolute or relative value. Using the labels from Table~\ref{tab:labels}, the agent evaluates alternative actions based on the hint text. For instance, for action~1, the agent generates the input text ``Set \verb|shared_buffers| to 25\% of RAM. \verb|shared_buffers| and 25\% relate to main memory.'', separating the two sentences via the type specification. If masked mode is activated, the two occurrences of the \verb|shared_buffers| parameter are masked. To make a choice, the agent internally compares values resulting from applying BERT to the input for each possible action.
\end{example}

\section{Aggregating Hints}
\label{sec:aggregate}

\begin{algorithm}[t]
\caption{Transition function for interpreting multiple hints.}
\label{alg:multistep}
\renewcommand{\algorithmiccomment}[1]{// #1}
\begin{algorithmic}[1]
\State \Comment{Given benchmark $b$, system properties $S$, multipli-}
\State \Comment{cators $M$, and weights $W$, translate and weigh hints $H$.}
\State \Comment{Evaluate $n$ configurations, created based on weighted hints.}
\Procedure{RunEpisode}{$b,H,S,M,W,n$}
\State $r_e\gets 0$
\State $H_w\gets\emptyset$
\State \Comment{Iterate over batch of hints}
\For{$\langle t,p,v\rangle\in H$}
\State \Comment{Translate hint into formula}
\State $d\gets 0$
\While{$d\in\{0, 1\}$}
\State $a\gets$\Call{ChooseAction}{$d,p,v,t$}
\State $\langle d,f,r_s\rangle\gets$\Call{TStep}{$-,p,v,S,M,d,f,a$}
\State $r_e\gets r_e+r_s$
\EndWhile
\State \Comment{Add weighted hint if admissible setting}
\If{$f\neq-$}
\State $a\gets$\Call{ChooseAction}{$d,p,v,t$}
\State $H_w\gets H_w\cup\{\langle W_a,p,f\rangle\}$
\EndIf
\EndFor
\State \Comment{Evaluate weighted hints in combination}
\State $r_e\gets r_e+$\Call{EvalWeighted}{$H_w,b,n$}
\State \Comment{Integrate new experiences}
\State \Call{UpdateRL}{$H_w,r_e$}
\EndProcedure
\end{algorithmic}
\end{algorithm}

The last section describes how to translate single tuning hints. However, we often need to integrate multiple hints, possibly from different sources, to obtain optimal performance. DB-BERT creates configurations based on groups of hints. This requires aggregating, possibly conflicting hints, from different sources. To support that, we expand the MDP presented in the last section. Instead of considering a single hint, we consider an entire batch of hints. For each single hint, we add an additional decision assigning the hint to a weight. This weight determines the priority when aggregating the hint with others into a configuration.

Algorithm~\ref{alg:multistep} shows complete pseudo-code executed during one iteration of DB-BERT's main loop (Algorithm~\ref{alg:multistep} is invoked by Algorithm~\ref{alg:main}). From the reinforcement learning perspective, each iteration corresponds to one episode of the associated MDP. Each episode starts from the same starting state, representing the default configuration. The number of hints considered per episode does therefore restrict the maximal number of changes, compared to the default configuration. However, as shown in recent work~\cite{VanAken2021, Kanellis2020}, tuning a small number of tuning knobs is typically sufficient to achieve near-optimal performance. 

Algorithm~\ref{alg:multistep} obtains a batch of candidate hints as input. It iterates over those hints and uses Algorithm~\ref{alg:singlestep} (Function~\textproc{Tstep}) to translate single hints (respectively, to determine that a candidate hint is erroneous and should not be considered). We postpone benchmark evaluations by specifying ``$-$'' as benchmark parameter for \textproc{Tstep}. If successful at translating the current hint into a formula (i.e., $f\neq -$), Algorithm~\ref{alg:multistep} assigns a weight (Line~18). Weights are chosen from a discrete set $W$ of possibilities and are assigned by the learning agent (Function~\textproc{ChooseAction}). Finally, the algorithm assembles a set $H_w$ of weighted tuning hints. 

\begin{algorithm}[t]
\caption{Evaluate set of weighted tuning hints on benchmark.}
\label{alg:evaluate}
\renewcommand{\algorithmiccomment}[1]{// #1}
\begin{algorithmic}[1]
\State \Comment{Maximal weighted distance from $V$ to nearest value in $C$.}
\Function{MaxDist}{$C,V$}
\State \Return{$\max_{\langle v,w\rangle\in V}w\cdot\min_{c\in C}\delta(v,c)$}
\EndFunction
\vspace{0.15cm}
\State \Comment{Evaluate performance of configuration $C$ on benchmark $b$.}
\Function{Evaluate}{$b,C$}
\State $suc\gets\mathrm{True}$
\For{$\langle p,v\rangle\in C$}
\State $suc\gets suc\wedge$\Call{DbmsSet}{$p,v$}
\EndFor
\If{$suc$}
\State \Return{\textproc{EvaluatePerformance}($b$)}
\Else
\State \Return{-1}
\EndIf
\EndFunction
\vspace{0.15cm}
\State \Comment{Evaluate up to $n$ configurations on benchmark $b$, }
\State \Comment{selecting configurations using weighted hints $H_w$.}
\Procedure{EvalWeighted}{$H_w,b,n$}
\State \Comment{Select configurations to cover hints}
\State $P\gets\{p|\langle w,p,v\rangle\}$
\State $C\gets\{\emptyset\}$
\For{$p\in P$}
\State $V\gets \{\langle v,w\rangle|w=\sum_{\langle w,p,v\rangle\in H_w}w\}$
\State $C_p\gets\emptyset$
\For{$i\gets1,\ldots,n$}
\State $v^*\gets\arg\min_{v|\langle v,w\rangle\in V}$\Call{MaxDist}{$C_p\cup\{v\},V$}
\State $C_p\gets C_p\cup\{v^*\}$
\EndFor
%\State $C\gets \{c\cup\{\langle p,v\rangle\}|c\in C,v\in C_p\}$
\EndFor
\State \Comment{Compose configurations to evaluate}
\State $C\gets\{\cup_{p\in P}i$-th entry from $C_p|1\leq i\leq n\}$
\State \Comment{Evaluate performance of configurations}
\State $E\gets\{$\Call{Evaluate}{$b,c$}$|c\in C\}$
\State \Return{$\max_{e\in E}e$}
\EndProcedure
\end{algorithmic}
\end{algorithm}

Next, we assemble one or several configurations to evaluate, using weighted hints. Algorithm~\ref{alg:evaluate} chooses and evaluates configurations, using weighted hints as input. It iterates over parameters mentioned in hints (loop from Line~23 to 30) and selects a limited number of $n$ values to try ($n$ is a tuning parameter). Values are selected in order to cover the range of suggested values (in hints) as well as possible. We choose values iteratively (loop from Line~26 to 29). We want to cover values proposed in hints as closely as possible in the following sense. Given a distance function $\delta$ comparing values for the same parameter, our goal is to minimize the maximal, weighted distance between a value proposed in a hint and the closest selected value. Function~\textproc{MaxDist} calculates the latter metric, given a weighted set $V$ of values and a set of selected configurations $C$. We select values greedily, minimizing the aforementioned cost function in each step\footnote{While this heuristic may seem simplistic, it can be shown that it finds near-optimal solutions. Consider the reduction of \textproc{MaxDist} as a function of selected values in $C$ (fixing $V$ and assigning \Call{MaxDist}{$\emptyset,V$} to a large constant). The reduction is sub-modular in the set of selected values, meaning that adding more values shows diminishing returns. As it is also non-negative and monotone (adding values cannot increase the maximal distance), the greedy algorithm corresponds to the algorithm by Nemhauser~\cite{Nemhauser1978} which guarantees solutions within factor $1-e^{-1}$ of the optimum.}. Note that some tuning knobs can only be set to specific values within their value range (e.g., MySQL's \verb|innodb_buffer_pool_size| must be a multiple of the chunk size~\cite{MySQL8manual}). We cannot simply average proposed values.

\begin{example}
Assume we collect hints recommending the following values for parameter \verb|shared_buffers|: 1~GB with weight 1, 2~GB with weight 8, and 8~GB with weight 1. When selecting 1~GB, we obtain maximal weighted distance of $8\cdot |2-1|=8$~GB from value 2~GB (only distance $1\cdot |8-1|=7$~GB from 8~GB). Selecting 2~GB yields a maximal weighted distance of 7~GB from value 8~GB. Selecting 8~GB yields a maximal weighted distance of 48~GB from value 2~GB. Hence, we select value 2~GB first. Next, we select value 8~GB to minimize the maximal distance to 1~GB.
\end{example}

Finally, we compose selected values for each parameter into $n$ configurations (Line~32). Function~\textproc{Evaluate} evaluates selected configurations on the given benchmark $b$. It assigns a penalty for configurations that are not accepted by the DBMS and, otherwise, calculates reward based on benchmark performance (we use the reward function introduced in Section~\ref{sub:environment}). Function~\textproc{EvalWeighted} returns the maximal reward obtained by any configuration.
\section{Experiments}
\label{sec:experiments}

We describe our experimental setup in Section~\ref{sub:setup}, provide details on the text documents used for NLP-enhanced database tuning in Section~\ref{sub:documents}, and details on the training process of all compared algorithms in Section~\ref{sub:training}. We compare DB-BERT against various baselines in Section~\ref{sub:baselines} and study the impact of text document size, data size, and various DB-BERT features on performance in Section~\ref{sub:further}.

% how to tune postgresql for tpc-h
% \url{http://rhaas.blogspot.com/2016/04/postgresql-96-with-parallel-query-vs.html}

% how to tune postgresl for tpc-c

\subsection{Experimental Setup}
\label{sub:setup}

We compare approaches for tuning system configuration parameters for MySQL~8.0 and for Postgres~13.2. We consider all numerical and Boolean tuning parameters that those systems offer: 232 parameters for Postgres and 266 parameters for MySQL. We use TPC-H with scaling factors one (Section~\ref{sub:baselines}) and ten (Section~\ref{sub:further}) and TPC-C with scaling factor 20 as benchmarks. For TPC-C, we use ten terminals, unlimited arrival rate, and 60 seconds for both, warmup and measurement time. Besides those parameters, we use the default TPC-C configurations for Postgres and MySQL from the OLTP benchmark\footnote{\url{https://github.com/oltpbenchmark/oltpbench}}. We execute five runs and allow for 25~minutes of tuning time (prior work uses the same time frame~\cite{Zhang2019a}). All experiments execute on a p3.2xlarge EC2 instance with 8~vCPUs, 61~GB of RAM, and a Tesla~V100 GPU featuring 16~GB of memory. The EC2 instance uses the Amazon Deep Learning AMI with Ubuntu~18.04. 

%For DB-BERT and other NLP-enhanced tuning methods, we did not observe significant improvements after 15~minutes of tuning. Hence, we reduce tuning time to 15~minutes when comparing NLP-enhanced tuning methods only.

%We allow 15~minutes of tuning time for each approach and benchmark and execute five runs. 

We compare against the recent DDPG++ algorithm~\cite{VanAken2021} as representative for tuning without NLP-enhancement. We consider different value ranges for tuning parameters, ranging from a factor of two around the default value (i.e., $d/2$ to $2\cdot d$ where $d$ is the default) to 100. We denote those versions as DDPG2, DDPG10, and DDPG100 in the following plots. Also, we compare against two baselines described in a recent vision paper on NLP-enhanced database tuning~\cite{Trummer2021}. In the following, Prior-Main denotes the main method proposed by that prior work, based on supervised learning. Also, we compare against a simple baseline, denoted as Prior-Simple, described in the same paper~\cite{Trummer2021}. 

By default, we use the following configuration parameters for DB-BERT. DB-BERT uses reinforcement learning to select multiplicator values and weights for each hint from a fixed set of alternatives. For all experiments, DB-BERT selects the multiplicator from the set $\{1/4,1/2,1,2,4\}$ and the weight from the set $\{0,2,4,8,16\}$. We use the same number of alternatives (five) in each case. This makes it easier to model the associated environment with OpenAI's gym framework. We avoid using overly small or large multiplicators (if the optimal parameter value deviates by more than factor four from the proposed value in any direction, the associated hint should be disregarded). The set of weight alternatives allows DB-BERT to disregard hints (by using a weight of zero) as well as to make specific hints up to eight times more important, compared to other hints with non-zero weights. We set $l$ to $10$ in order to allow at most ten hints per episode and parameter. We consider at most 50 hints per episode in total ($e=50$) and evaluate two configurations per episode ($n=2$). DB-BERT splits text documents into segments of length at most 128 tokens.

%As multiplicators $M$ (used to deviate from concrete parameter values proposed in tuning hints), we use powers of two, ranging from $2^{-2}$ to $2^2$. As weights $W$ (used to determine relative importance of conflicting hints), we use powers of two from $2^1$ to $2^4$ and the zero weight. \rev{}

All baselines are implemented in Python~3.7, using Pytorch~1.8.1 and (for the NLP-enhanced tuning baselines) the Huggingface Transformers library~\cite{Wolf2020}. DB-BERT uses Google's programmable search engine API\footnote{\url{https://developers.google.com/custom-search}} to retrieve text documents. Also, DB-BERT uses the Double Deep Q-Networks~\cite{VanHasselt2016} implementation from the Autonomous Learning Library\footnote{\url{https://github.com/cpnota/autonomous-learning-library}} as reinforcement learning algorithm.

\subsection{Tuning Text Documents}
\label{sub:documents}

\begin{figure}
    \centering
    \begin{tikzpicture}
        \begin{groupplot}[group style={group size=2 by 2}, width=4.5cm, height=3cm, xmajorgrids, ylabel near ticks, no markers]
            \nextgroupplot[title={Postgres}, xlabel=Parameters, ylabel={Nr.\ Docs}]
            \addplot table[x index=0, y index=2, col sep=tab, header=false] {plots/doc_analysis/pg_param};
            \nextgroupplot[title={MySQL}, xlabel=Parameters, ylabel={Nr.\ Docs}]
            \addplot table[x index=0, y index=2, col sep=tab, header=false] {plots/doc_analysis/ms_param};
            \nextgroupplot[xlabel=Hints, ylabel={Nr.\ Occurrences}]
            \addplot table[x index=0, y index=2, col sep=tab, header=false] {plots/doc_analysis/pg_asg};
            \nextgroupplot[xlabel=Hints, ylabel={Nr.\ Occurrences}]
            \addplot table[x index=0, y index=2, col sep=tab, header=false] {plots/doc_analysis/ms_asg};
        \end{groupplot}
    \end{tikzpicture}
    \caption{Frequency distribution of hints and parameters in the collection of tuning documents for Postgres and MySQL.}
    \label{fig:docanalysis}
\end{figure}
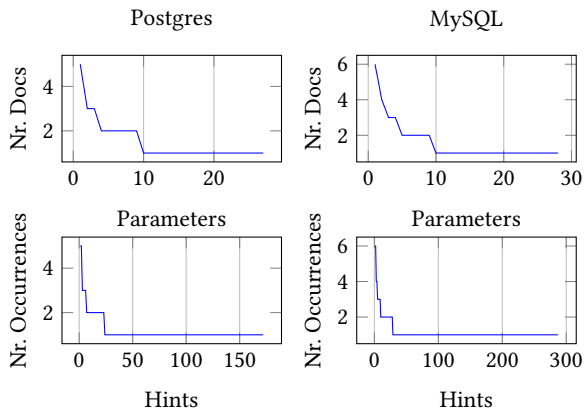

DB-BERT comes with a script that retrieves text documents via Google search and transforms them into the input format required by DB-BERT. For most of the following experiments, we use two document collections retrieved via the queries ``Postgresql performance tuning hints'' (issued on April 11, 2021) and ``MySQL performance tuning hints'' (issued on April 15, 2021). We included the first 100 Google results for each of the two queries into the corresponding document collection (accounting for a total of 1.3~MB of text for Postgres and 2.4~MB of text for MySQL). The results are diverse and cover blog entries, forum discussions (e.g., on Database Administrators Stack Exchange\footnote{\url{https://dba.stackexchange.com/}}), as well as the online manuals from both database systems. We call the document collection for Postgres Pg100 and the one for MySQL Ms100 in the following. 

Figure~\ref{fig:docanalysis} shows the distribution of parameter mentions and proposed value assignments in those document collections, generated via DB-BERT's candidate hint extraction mechanism (see Section~\ref{sec:extract}). Clearly, the distribution of hints over documents and parameters is non-uniform. For both database systems, few parameters are mentioned in multiple documents while most parameters are mentioned only in a single document. Similarly, there are a few assignments proposed by multiple sources. On the other side, most value assignments are proposed only once.

\begin{table}
    \centering
    \caption{Tuning parameters mentioned in most documents for Postgres and MySQL.}
    \begin{tabular}{ll}
        \toprule[1pt]
         \textbf{System} & \textbf{Parameter} \\
         \midrule[1pt]
         Postgres & \verb|shared_buffers| \\
         & \verb|max_connections| \\
         & \verb|max_parallel_workers_per_gather| \\
         & \verb|max_wal_size| \\
         & \verb|wal_buffers| \\
         \midrule[1pt]
         MySQL & \verb|innodb_buffer_pool_size| \\
         & \verb|join_buffer_size| \\
         & \verb|innodb_buffer_pool_instances| \\
         & \verb|max_connections| \\
         & \verb|innodb_flush_log_at_trx_commit| \\
         \bottomrule[1pt]
    \end{tabular}
    \label{tab:frequentparams}
\end{table}

Table~\ref{tab:frequentparams} shows the most frequently mentioned parameters for both Postgres and MySQL. Parameters related to buffer size (e.g., \verb|shared_buffers| for Postgres and \verb|innodb_buffer_pool_size| for MySQL) feature prominently among them. Besides that, parameters related to parallelism (e.g.,  \verb|max_parallel_workers_per_gather|) or logging (e.g., \verb|max_wal_size|) are mentioned frequently as well. 

% and \verb|innodb_flush_log_at_trx_commit|
%\subsection{Training Phase}

\subsection{Training}
\label{sub:training}

Two of the compared algorithms, namely DB-BERT and Prior-Main, use training before run time. Prior-Main uses natural language tuning hints, annotated with associated formulas, as training data. We use the same training samples and training parameters as in the prior work~\cite{Trummer2021}. Consistent with the experimental setup in the latter paper, we apply Prior-Main, trained on Postgres samples, to tune MySQL and Prior-Main, trained on MySQL samples, to tune Postgres. The goal is to demonstrate that NLP-enhanced database tuning does not require system-specific, annotated samples.

Prior-Main has no support for extracting benchmark-specific tuning hints from a fixed document collection, a disadvantage if the same document collection is used for tuning multiple benchmarks. To allow at least some degree of variability, we train the Prior-Main model separately for each of our five benchmark runs. This leads to slightly different extractions in each run. Training Prior-Main on the platform outlined in Section~\ref{sub:setup} took 417 seconds for MySQL samples and 393 for Postgres samples.

DB-BERT does not use annotated tuning hints for training. Instead, it uses the database system itself for run time feedback during the training phase. Similar to Prior-Main, we train DB-BERT on Pg100 to tune MySQL and on Ms100 to tune Postgres. We activate the masked mode during training (see Section~\ref{sec:translate}), meaning that parameter names are masked. This avoids learning system-specific parameter names (which are useless in our experimental setup) and focuses attention on the sentence structure of tuning hints instead. The reward signal of DB-BERT (see Section~\ref{sec:translate} and \ref{sec:aggregate}) combines reward for successfully changing parameter values according to tuning hints (meaning that the corresponding values are valid) and for performance obtained. To measure performance, we use a synthetic database containing two tables with two columns containing consecutive numbers from 1 to 1,000,000. We use a simple count aggregation query joining both tables with an equality predicate. Reward for performance is scaled down by a factor of 100 to avoid specialization to this artificial benchmark (it merely serves to penalize particularly bad configurations such as setting the buffer pool size to a minimal value). Finally, we add a small reward bonus for setting parameter values that are within the same order of magnitude as the default setting (assuming that extreme deviations from default values are possible but less likely). DB-BERT's training starts from the BERT base model~\cite{Devlin2019a} with 110~million parameters. All model parameters are tuned during training. 

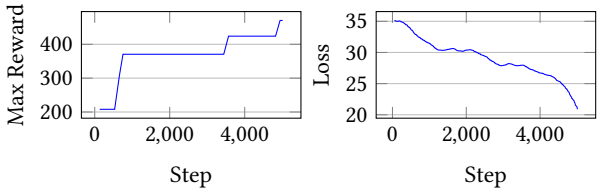
\begin{figure}
    \centering
    \begin{tikzpicture}
        \begin{groupplot}[group style={group size=3 by 1}, width=4.5cm, height=3cm, xlabel=Step, ymajorgrids, ylabel near ticks, no markers]
            \nextgroupplot[ylabel=Max Reward]
            \addplot table[x=Step, y=Value, col sep=comma] {plots/training/pg100maxreturns.csv};
            \nextgroupplot[ylabel=Loss]
            \addplot table[x=Step, y=Smoothed, col sep=comma] {plots/training/pg100loss.csv};
        \end{groupplot}
    \end{tikzpicture}
    \caption{Reward and loss when training DB-BERT on Pg100.}
    \label{fig:training}
\end{figure}

We trained DB-BERT for 5,000 iterations on Pg100 and for 10,000 iterations on Ms100 (due to the larger number of hints in this collection). Training took 43 minutes for Pg100 and 84 minutes for Ms100. Figure~\ref{fig:training} shows progress for Pg100 as a function of the number of training steps.

%In summary, DB-BERT is incentivized to learn translating tuning hints into admissible configuration changes that do not degrade performance and are not extremely different from default settings.

\subsection{Comparison with Baselines}
\label{sub:baselines}

\begin{figure}
    \centering
    \begin{tikzpicture}
        \begin{groupplot}[group style={group size=1 by 2, ylabels at=edge left, xlabels at=edge bottom}, width=8cm, height=4.8cm, legend entries={DDPG2, DDPG10, DDPG100, DB-BERT, Prior-Simple, Prior-Main}, legend columns=3, ymode=normal, ymajorgrids, xlabel={Optimization Time (s)}, ylabel={Execution Time (s)}, legend to name=ddpgTpchLegend, cycle list name={baselinescyclelist}]
            \nextgroupplot[title=Postgres, mark size=0.5]
            \addplot+[error bars/.cd, y dir=both, y explicit] table[x expr=\thisrow{millis}*0.001, y expr=\thisrow{avg}*0.001, y error minus expr=\thisrow{avg}*0.001-\thisrow{p20}*0.001, y  error plus expr=\thisrow{p80}*0.001-\thisrow{avg}*0.001, header=true, col sep=comma] {plots/25min/ddpg/pg_tpch_t2_plot};
            \addplot+[error bars/.cd, y dir=both, y explicit] table[x expr=\thisrow{millis}*0.001, y expr=\thisrow{avg}*0.001, y error minus expr=\thisrow{avg}*0.001-\thisrow{p20}*0.001, y  error plus expr=\thisrow{p80}*0.001-\thisrow{avg}*0.001, header=true, col sep=comma] {plots/25min/ddpg/pg_tpch_t10_plot};
            \addplot+[error bars/.cd, y dir=both, y explicit] table[x expr=\thisrow{millis}*0.001, y expr=\thisrow{avg}*0.001, y error minus expr=\thisrow{avg}*0.001-\thisrow{p20}*0.001, y  error plus expr=\thisrow{p80}*0.001-\thisrow{avg}*0.001, header=true, col sep=comma] {plots/25min/ddpg/pg_tpch_t100_plot};
            \addplot+[error bars/.cd, y dir=both, y explicit] table[x expr=\thisrow{millis}*0.001, y expr=\thisrow{avg}*0.001, y error minus expr=\thisrow{avg}*0.001-\thisrow{p20}*0.001, y  error plus expr=\thisrow{p80}*0.001-\thisrow{avg}*0.001, header=true, col sep=comma] {plots/25min/dbbert/pg_tpch_base_plot};
           \addplot+[error bars/.cd, y dir=both, y explicit] coordinates {(23, 22) +- (0, 0.5)};
            \addplot+[error bars/.cd, y dir=both, y explicit] coordinates {(515, 20) += (0, 2.4) -= (0,3.6)};
            
           \nextgroupplot[title=MySQL, mark size=0.5]
            \addplot+[error bars/.cd, y dir=both, y explicit] table[x expr=\thisrow{millis}*0.001, y expr=\thisrow{avg}*0.001, y error minus expr=\thisrow{avg}*0.001-\thisrow{p20}*0.001, y  error plus expr=\thisrow{p80}*0.001-\thisrow{avg}*0.001, header=true, col sep=comma] {plots/25min/ddpg/ms_tpch_t2_plot};
            \addplot+[error bars/.cd, y dir=both, y explicit] table[x expr=\thisrow{millis}*0.001, y expr=\thisrow{avg}*0.001, y error minus expr=\thisrow{avg}*0.001-\thisrow{p20}*0.001, y  error plus expr=\thisrow{p80}*0.001-\thisrow{avg}*0.001, header=true, col sep=comma] {plots/25min/ddpg/ms_tpch_t10_plot};
            \addplot+[error bars/.cd, y dir=both, y explicit] table[x expr=\thisrow{millis}*0.001, y expr=\thisrow{avg}*0.001, y error minus expr=\thisrow{avg}*0.001-\thisrow{p20}*0.001, y  error plus expr=\thisrow{p80}*0.001-\thisrow{avg}*0.001, header=true, col sep=comma] {plots/25min/ddpg/ms_tpch_t100_plot};
            \addplot+[error bars/.cd, y dir=both, y explicit] table[x expr=\thisrow{millis}*0.001, y expr=\thisrow{avg}*0.001, y error minus expr=\thisrow{avg}*0.001-\thisrow{p20}*0.001, y  error plus expr=\thisrow{p80}*0.001-\thisrow{avg}*0.001, header=true, col sep=comma] {plots/25min/dbbert/ms_tpch_base_plot};
            \addplot+[error bars/.cd, y dir=both, y explicit] coordinates {(137, 136) += (0,1) -= (0, 1)};
            \addplot+[error bars/.cd, y dir=both, y explicit] coordinates {(845, 114) += (0,23) -= (0, 12)};
        \end{groupplot}    
    \end{tikzpicture}
    
    \ref{ddpgTpchLegend}
    \caption{Minimal execution time for TPC-H as a function of optimization time for different baselines.}
    \label{fig:ddpgTpch}
\end{figure}
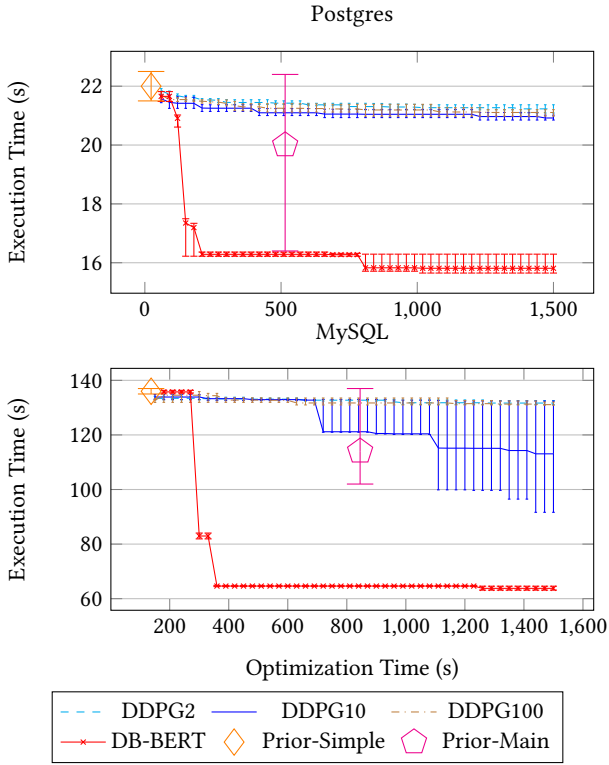

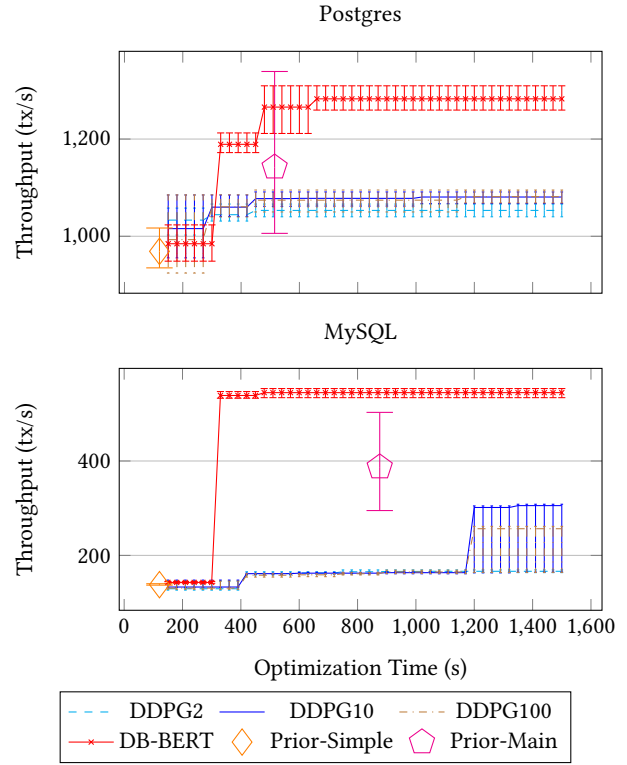
\begin{figure}
    \centering
    \begin{tikzpicture}
        \begin{groupplot}[group style={group size=1 by 2, ylabels at=edge left, x descriptions at=edge bottom}, width=8cm, height=4.8cm, legend entries={DDPG2, DDPG10, DDPG100, DB-BERT, Prior-Simple, Prior-Main}, legend columns=3, ymode=normal, ymajorgrids, xlabel={Optimization Time (s)}, ylabel={Throughput (tx/s)}, legend to name=ddpgTpccLegend, cycle list name={baselinescyclelist}]
            \nextgroupplot[title=Postgres, mark size=0.5]
            % Raw input data runs TPCC transactions for 60 seconds but calculates averages for bins of 120 seconds -> need to multiply by factor 2
            \addplot+[error bars/.cd, y dir=both, y explicit] table[x expr=\thisrow{millis}*0.001, y expr=\thisrow{avg}*2, y error plus expr=\thisrow{p80}*2-\thisrow{avg}*2, y error minus expr=\thisrow{avg}*2-\thisrow{p20}*2,  header=true, col sep=comma] {plots/25min/ddpg/pg_tpcc_t2_plot};
            \addplot+[error bars/.cd, y dir=both, y explicit] table[x expr=\thisrow{millis}*0.001, y expr=\thisrow{avg}*2, y error plus expr=\thisrow{p80}*2-\thisrow{avg}*2, y error minus expr=\thisrow{avg}*2-\thisrow{p20}*2,  header=true, col sep=comma] {plots/25min/ddpg/pg_tpcc_t10_plot};
            \addplot+[error bars/.cd, y dir=both, y explicit] table[x expr=\thisrow{millis}*0.001, y expr=\thisrow{avg}*2, y error plus expr=\thisrow{p80}*2-\thisrow{avg}*2, y error minus expr=\thisrow{avg}*2-\thisrow{p20}*2,  header=true, col sep=comma] {plots/25min/ddpg/pg_tpcc_t100_plot};
            \addplot+[error bars/.cd, y dir=both, y explicit] table[x expr=\thisrow{millis}*0.001, y expr=\thisrow{avg}*2, y error plus expr=\thisrow{p80}*2-\thisrow{avg}*2, y error minus expr=\thisrow{avg}*2-\thisrow{p20}*2,  header=true, col sep=comma] {plots/25min/dbbert/pg_tpcc_base_plot};
           \addplot+[error bars/.cd, y dir=both, y explicit] coordinates {(120, 969) += (0, 48) -= (0,34)};
           \addplot+[error bars/.cd, y dir=both, y explicit] coordinates {(515, 1142) += (0, 197) -= (0,136)};
            
           \nextgroupplot[title=MySQL, mark size=0.5]
            \addplot+[error bars/.cd, y dir=both, y explicit] table[x expr=\thisrow{millis}*0.001, y expr=\thisrow{avg}*2, y error plus expr=\thisrow{p80}*2-\thisrow{avg}*2, y error minus expr=\thisrow{avg}*2-\thisrow{p20}*2,  header=true, col sep=comma] {plots/25min/ddpg/ms_tpcc_t2_plot};
            \addplot+[error bars/.cd, y dir=both, y explicit] table[x expr=\thisrow{millis}*0.001, y expr=\thisrow{avg}*2, y error plus expr=\thisrow{p80}*2-\thisrow{avg}*2, y error minus expr=\thisrow{avg}*2-\thisrow{p20}*2,  header=true, col sep=comma] {plots/25min/ddpg/ms_tpcc_t10_plot};
            \addplot+[error bars/.cd, y dir=both, y explicit] table[x expr=\thisrow{millis}*0.001, y expr=\thisrow{avg}*2, y error plus expr=\thisrow{p80}*2-\thisrow{avg}*2, y error minus expr=\thisrow{avg}*2-\thisrow{p20}*2,  header=true, col sep=comma] {plots/25min/ddpg/ms_tpcc_t100_plot};
            \addplot+[error bars/.cd, y dir=both, y explicit] table[x expr=\thisrow{millis}*0.001, y expr=\thisrow{avg}*2, y error plus expr=\thisrow{p80}*2-\thisrow{avg}*2, y error minus expr=\thisrow{avg}*2-\thisrow{p20}*2,  header=true, col sep=comma] {plots/25min/dbbert/ms_tpcc_base_plot};
            \addplot+[error bars/.cd, y dir=both, y explicit] coordinates {(121, 138) += (0, 2) -= (0,1)};
            \addplot+[error bars/.cd, y dir=both, y explicit] coordinates {(876, 387) += (0,116) -= (0, 92)};
        \end{groupplot}
    \end{tikzpicture}
    
    \ref{ddpgTpccLegend}
    \caption{Maximal throughput for TPC-C as a function of optimization time for different baselines.}
    \label{fig:ddpgTpcc}
\end{figure}

We compare DB-BERT against baselines on TPC-H (see Figure~\ref{fig:ddpgTpch}) and TPC-C (see Figure~\ref{fig:ddpgTpcc}). We tune Postgres and MySQL for 25 minutes per run. We use throughput as optimization metric for TPC-C and execution time for TPC-H. We show performance of the best configuration found (y axis) as a function of optimization time (x axis). In these and the following plots, we report the arithmetic average as well as the 20th and 80th percentile of five runs (using error bars to show percentiles). 

% (e.g., seven to eight iterations for TPC-C)

DDPG++~\cite{VanAken2021} is a database tuning approach, based on reinforcement learning. It was shown to be competitive with various other state of the art tuning approaches~\cite{VanAken2021}. However, the prior publication evaluates DDPG++ for a few tens of tuning parameters and allocates 150 iterations per tuning session. Here, we consider hundreds of parameters for tuning and aim at a tuning time frame that allows only few iterations. Clearly, within the allocated time frame, DDPG++ does not find solutions of comparable quality to DB-BERT. In particular for TPC-H, DDPG++ often tries parameter changes that decrease performance significantly (e.g., changes to optimizer cost constants triggering different join orders). Hence, performance of the best configuration found remains almost constant for DDPG++ (close to the one achieved via the default configuration, tried before the initial iteration). DDPG++ could benefit from specifying parameter-specific value ranges to consider during tuning. For instance, increasing buffer pool size by an order of magnitude, compared to the default settings, is often beneficial. For optimizer cost constants (e.g., \verb|random_page_cost| in Postgres), doing so is however dangerous. Our goal is to show that such input can be partially substituted by information mined automatically from text.

% that lead to either inadmissible changes parameter settings with detrimental impact on performance randomly changing optimizer cost constants, for instance, tends to have detrimental impact on performance. Hence, 

% DDPG++ typically cannot improve over the default configuration (evaluated at the start of tuning), in particular due to parameters that  Of course, DDPG++ has been designed for a different optimization time frame and uses less information as input. Our goal is merely to show that additional information, presented in the form of tuning text, can help to focus tuning efforts and reduce tuning time.

%This does not mean that DDPG++ is an inferior approach in general. 

Prior-Simple and Prior-Main are the two most related baselines as both use tuning text as input, similar to DB-BERT. Prior-Simple uses a naive heuristic for translation. Applying this heuristic is fast and Prior-Simple is typically the first baseline to return results. However, it only extracts the recommendation to set \verb|checkpoint_completion_target| to 0.9 from Pg100 and no recommendations from Ms100. Hence, it does not improve over the default configuration. Prior-Main performs significantly better. Due to small differences in training, extractions differ across different runs, leading to high variance. For instance, for Pg100, Prior-Main is able to extract a tuning hint that recommends setting \verb|shared_buffers| to 25\% of main memory in two out of five runs. This can lead to significant performance improvements, in particular for TPC-H. However, average performance is significantly below the optimum. As Prior-Main classifies all sentences in the document collection before aggregating tuning hints, its run time is significantly higher than the one of Prior-Simple.

DB-BERT achieves attractive tradeoffs between tuning time and result quality. Unlike DDPG++, it uses tuning text as input that allows identifying the most relevant parameters and candidate values quickly. Compared to Prior-Simple and Prior-Main, it finds significantly better solutions in average. In particular for MySQL, Prior-Main typically fails to find solutions of comparable quality. Furthermore, the time taken by Prior-Main to analyze all documents is typically higher by a factor of two to three, compared to the time until DB-BERT produces a near-optimal solution (i.e., within one percent of DB-BERT's final optimum). Tables~\ref{tab:pg_tpch_configuration} and \ref{tab:pg_tpcc_configuration} show configurations found by DB-BERT when tuning Postgres. Despite extracting hints from the same document collection, DB-BERT is able to find benchmark-specific configurations.

%This shows the benefit of hint ordering. 

% For theses and the following plots, we report the arithmetic average

% Simple baseline on EC2 p3.2xlarge server. postgres100 - takes 1 second. Recommends \verb|checkpoint_completion_target|, 0.9. mysql100 takes 1.3 seconds. No recommendations. On job file: 0.5 seconds, no recommendations.

% Prior supervised learning on EC2 p3.x2large server. Training on mysql100 takes 6m57s, for pg100 it takes 6m33s. Inference on postgres100 (trained from ms100): 8m33s. Recommends setting \verb|shared_buffers| to $25\%$. Inference on mysql100 (trained from pg100) takes 12m36s. At least two sources recommend setting \verb|innodb_buffer_pool_instances| to 1, \verb|innodb_buffer_pool_size| to 1GB, \verb|innodb_buffer_pool_instances| to 1GB, further setting \verb|innodb_buffer_pool_size| to 2, and \verb|innodb_buffer_pool_size| to 3. Always set restrictive parameter to ``False''. Finally, using inference on \verb|pg_job_single| from \verb|pg_10| takes 11 seconds, returns no recommendation (not even from one source). Same when using model trained on \verb|ms_10|.

% Using VLDB'21 recommendations for MySQL. TPC-H, run time in seconds:
% 80.4 average, error is 1 second. TPC-C throughput: 611 average, error 27

%\pgfplotscreateplotcyclelist{baselinescyclelist}{{cyan, mark=o, mark size=2},{blue, mark=triangle, mark size=2},{brown, mark=square, mark size=2},{red, mark=x, mark size=2},{orange, mark=diamond, mark size=5, only marks},{magenta, mark=pentagon, mark size=5, only marks},{gray, mark=otimes*, mark size=5, only marks}}

\pgfplotscreateplotcyclelist{baselinescyclelist}{{cyan, dashed},{blue, solid},{brown, dashdotted},{red, mark=x, mark size=1.35},{orange, mark=diamond, mark size=5, only marks},{magenta, mark=pentagon, mark size=5, only marks},{gray, mark=otimes*, mark size=5, only marks}}

\begin{table}[t]
    \centering
    \caption{Postgres configuration for TPC-H by DB-BERT.}
    \begin{tabular}{ll}
    \toprule[1pt]
    \textbf{Parameter} & \textbf{Value} \\
    \midrule[1pt]
         \verb|max_connections| & 1100 \\
         \verb|max_parallel_workers_per_gather| & 19 \\
         \verb|max_wal_size| & 4GB \\
         \verb|shared_buffers| & 1GB \\
         \bottomrule[1pt]
    \end{tabular}
    \label{tab:pg_tpch_configuration}
\end{table}

\begin{table}[t]
    \centering
    \caption{Postgres configuration for TPC-C by DB-BERT.}
    \begin{tabular}{ll}
    \toprule[1pt]
    \textbf{Parameter} & \textbf{Value} \\
    \midrule[1pt]
         \verb|archive_command| & 3 \\
         \verb|archive_timeout| & 4 \\
         \verb|checkpoint_flush_after| & 0 \\
         \verb|maintenance_work_mem| & 32MB \\
         \verb|max_wal_senders| & 5 \\
         \verb|random_page_cost| & 2 \\
         \verb|synchronous_commit| & 0 \\
         \bottomrule[1pt]
    \end{tabular}
    \label{tab:pg_tpcc_configuration}
\end{table}

\begin{comment}
\begin{table}[t]
    \centering
    \begin{tabular}{ll}
    \toprule[1pt]
    \textbf{Parameter} & \textbf{Value} \\
    \midrule[1pt]
         \verb|innodb_buffer_pool_size| & 8GB \\
         \verb|join_buffer_size| & 2MB \\
         \verb|max_connections| & 3200 \\
         \bottomrule[1pt]
    \end{tabular}
    \caption{DB-BERT's best configuration for MySQL.}
    \label{tab:ms_configuration}
\end{table}
\end{comment}

\subsection{Further Analysis}
\label{sub:further}

\begin{figure}
    \centering
    \begin{tikzpicture}
        \begin{groupplot}[group style={group size=1 by 1, ylabels at=edge left, xlabels at=edge bottom}, width=8cm, height=4.8cm, legend entries={DB-BERT, No Hint Reordering, No Implicit Hints}, legend columns=4, ymode=normal, ymajorgrids, xlabel={Optimization Time (s)}, ylabel={Execution Time (s)}, legend to name=variantsLegend, cycle list={{red, mark=x, mark size=1.35}, {blue, mark=o, mark size=2}, {}}]
            \nextgroupplot[title=Postgres, mark size=0.5]
            \addplot+[error bars/.cd, y dir=both, y explicit] table[x expr=\thisrow{millis}*0.001, y expr=\thisrow{avg}*0.001, y error minus expr=\thisrow{avg}*0.001-\thisrow{p20}*0.001, y  error plus expr=\thisrow{p80}*0.001-\thisrow{avg}*0.001, header=true, col sep=comma] {plots/25min/dbbert/pg_tpch_base_plot};
            \addplot+[error bars/.cd, y dir=both, y explicit] table[x expr=\thisrow{millis}*0.001, y expr=\thisrow{avg}*0.001, y error minus expr=\thisrow{avg}*0.001-\thisrow{p20}*0.001, y  error plus expr=\thisrow{p80}*0.001-\thisrow{avg}*0.001, header=true, col sep=comma] {plots/25min/variants/pg_tpch_by_doc_25min_performance_plot};
            \addplot+[error bars/.cd, y dir=both, y explicit] table[x expr=\thisrow{millis}*0.001, y expr=\thisrow{avg}*0.001, y error minus expr=\thisrow{avg}*0.001-\thisrow{p20}*0.001, y  error plus expr=\thisrow{p80}*0.001-\thisrow{avg}*0.001, header=true, col sep=comma] {plots/25min/variants/pg_tpch_no_implicit_25min_performance_plot};
        \end{groupplot}    
    \end{tikzpicture}
    
    \ref{variantsLegend}
    \caption{Comparison of different DB-BERT variants when optimizing Postgres for TPC-H.}
    \label{fig:variants}
\end{figure}
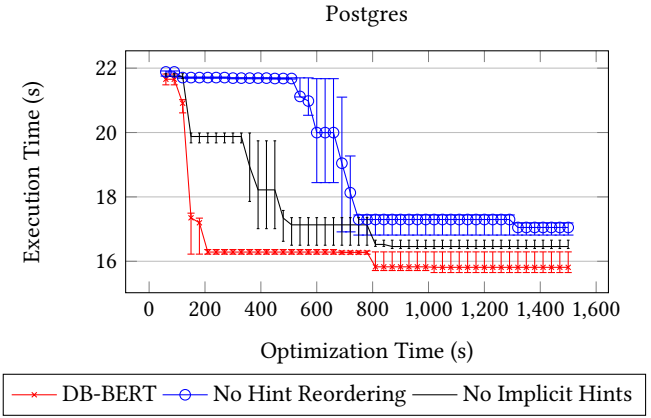

We study the impact of different factors on tuning performance. First, we compare DB-BERT against two simplified variants in Figure~\ref{fig:variants}. We compare against a variant of DB-BERT that processes hints in document order (instead of prioritizing them as described in Section~\ref{sec:extract}). Also, we compare against a variant that does not consider implicit hints (i.e., only hints where parameter names are explicitly mentioned). Clearly, both simplifications degrade tuning performance on TPC-H. Considering hints in document order prevents DB-BERT from tuning the most relevant parameters first. Discarding implicit hints reduces the total set of available hints. 

\begin{figure}
    \centering
    \begin{tikzpicture}
        \begin{groupplot}[group style={group size=1 by 1, ylabels at=edge left, xlabels at=edge bottom}, width=8cm, height=4.8cm, legend entries={DB-BERT: Generic, DB-BERT: Specific, Prior-Simple: Generic, Prior-Main: Generic, Prior-Simple: Specific, Prior-Main: Specific}, legend columns=2, ymode=normal, ymajorgrids, xlabel={Optimization Time (s)}, ylabel={Execution Time (s)}, legend to name=documentsLegend, cycle list={{red, mark=x, mark size=1.35},{blue, mark=o, mark size=2},{orange, mark=diamond, mark size=5, only marks},{magenta, mark=pentagon, mark size=5, only marks},{blue, mark=square, mark size=5},{blue, mark=triangle, mark size=5}}]
            \nextgroupplot[title=Postgres, mark size=0.5]
            \addplot+[error bars/.cd, y dir=both, y explicit] table[x expr=\thisrow{millis}*0.001, y expr=\thisrow{avg}*0.001, y error minus expr=\thisrow{avg}*0.001-\thisrow{p20}*0.001, y  error plus expr=\thisrow{p80}*0.001-\thisrow{avg}*0.001, header=true, col sep=comma] {plots/25min/dbbert/pg_tpch_base_plot};
            \addplot+[error bars/.cd, y dir=both, y explicit] table[x expr=\thisrow{millis}*0.001, y expr=\thisrow{avg}*0.001, y error minus expr=\thisrow{avg}*0.001-\thisrow{p20}*0.001, y  error plus expr=\thisrow{p80}*0.001-\thisrow{avg}*0.001, header=true, col sep=comma] {plots/25min/variants/pg_tpch_small_25min_performance_plot};

            \addplot+[only marks, error bars/.cd, y dir=both, y explicit] coordinates {(23, 22) +- (0, 0.5)};
            \addplot+[only marks, error bars/.cd, y dir=both, y explicit] coordinates {(515, 20) += (0, 2.4) -= (0,3.6)};
            \addplot+[only marks, error bars/.cd, y dir=both, y explicit] coordinates {(23, 22) += (0, 0.1) -= (0, 0.05)};
            \addplot+[only marks, error bars/.cd, y dir=both, y explicit] coordinates {(44, 21.9) += (0, 0.1) -= (0, 0)};

        \end{groupplot}    
    \end{tikzpicture}
    
    \ref{documentsLegend}
    \caption{NLP-enhanced database tuning for TPC-H on Postgres with different input text (100 documents with generic hints versus one document with benchmark-specific hints).}
    \label{fig:documents}
\end{figure}
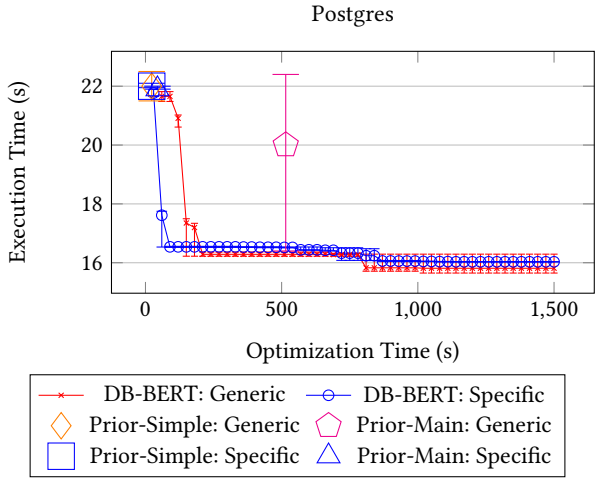

Next, we study the impact of the input text. We replace Pg100, containing hundreds of generic tuning hints, by a single blog post\footnote{\url{http://rhaas.blogspot.com/2016/04/postgresql-96-with-parallel-query-vs.html}}. This post describes how to tune Postgres specifically for TPC-H. Figure~\ref{fig:docanalysis} compares performance with different input documents for all NLP-enhanced tuning baselines. While the performance of Prior-Simple does not change with the input text, the performance of Prior-Main degrades as we switch to the smaller document. Prior-Main benefits from large document collections as redundant hints can partially make up for imprecise extractions. For the smaller input document, it does not extract any hints. DB-BERT, however, benefits from more specialized tuning hints. Using benchmark-specific input text, it converges to near-optimal solutions faster and ultimately finds a slightly better solution (using a higher value for the \verb|shared_buffers| parameter, compared to Table~\ref{tab:pg_tpch_configuration}, as proposed in the blog entry). 

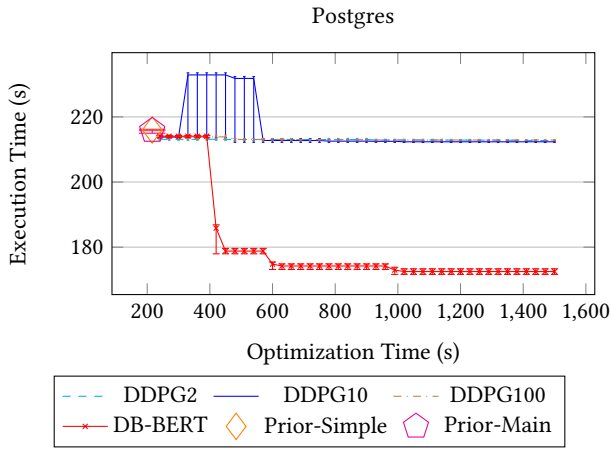
\begin{figure}
    \centering
    \begin{tikzpicture}
        \begin{groupplot}[group style={group size=1 by 1, ylabels at=edge left, xlabels at=edge bottom}, width=8cm, height=4.8cm, legend entries={DDPG2, DDPG10, DDPG100, DB-BERT, Prior-Simple, Prior-Main}, legend columns=3, ymode=normal, ymajorgrids, xlabel={Optimization Time (s)}, ylabel={Execution Time (s)}, legend to name=largeTpchLegend, cycle list name={baselinescyclelist}]
            \nextgroupplot[title=Postgres, mark size=0.5]
            \addplot+[error bars/.cd, y dir=both, y explicit] table[x expr=\thisrow{millis}*0.001, y expr=\thisrow{avg}*0.001, y error minus expr=\thisrow{avg}*0.001-\thisrow{p20}*0.001, y  error plus expr=\thisrow{p80}*0.001-\thisrow{avg}*0.001, header=true, col sep=comma] {plots/25min/large/pg_tpch_t2_25min_performance_plot};
            \addplot+[error bars/.cd, y dir=both, y explicit] table[x expr=\thisrow{millis}*0.001, y expr=\thisrow{avg}*0.001, y error minus expr=\thisrow{avg}*0.001-\thisrow{p20}*0.001, y  error plus expr=\thisrow{p80}*0.001-\thisrow{avg}*0.001, header=true, col sep=comma] {plots/25min/large/pg_tpch_t10_25min_performance_plot};
            \addplot+[error bars/.cd, y dir=both, y explicit] table[x expr=\thisrow{millis}*0.001, y expr=\thisrow{avg}*0.001, y error minus expr=\thisrow{avg}*0.001-\thisrow{p20}*0.001, y  error plus expr=\thisrow{p80}*0.001-\thisrow{avg}*0.001, header=true, col sep=comma] {plots/25min/large/pg_tpch_t100_25min_performance_plot};
            \addplot+[error bars/.cd, y dir=both, y explicit] table[x expr=\thisrow{millis}*0.001, y expr=\thisrow{avg}*0.001, y error minus expr=\thisrow{avg}*0.001-\thisrow{p20}*0.001, y  error plus expr=\thisrow{p80}*0.001-\thisrow{avg}*0.001, header=true, col sep=comma] {plots/25min/large/dbbert_25min_performance_plot};
            
        \addplot+[error bars/.cd, y dir=both, y explicit] coordinates {(216, 216) +- (0, 0.25)};
        \addplot+[error bars/.cd, y dir=both, y explicit] coordinates {(216, 215.8) += (0, 0.2) -= (0,0.8)};
        \end{groupplot}    
    \end{tikzpicture}
    
    \ref{largeTpchLegend}
    \caption{Minimal execution time for TPC-H with scaling factor 10 as a function of optimization time.}
    \label{fig:largeTpch}
\end{figure}

Finally, we scale up the data size. Figure~\ref{fig:largeTpch} reports results for TPC-H with scaling factor 10 (and using the TPC-H specific tuning text)\footnote{Note that execution time for the best configuration increases slightly at the beginning for DDPG10. This cannot happen as long as we consider a single run. However, we average over a smaller set of runs that finished their first evaluation fast for the first data point, while the second data point averages over all runs.}. Compared to Figure~\ref{fig:documents}, showing results for scaling factor one, it takes longer for DB-BERT to find near-optimal solutions. This is expected, as longer run times per benchmark evaluation reduce the number of DB-BERT's iterations per time unit. Compared to other baselines, DB-BERT finds significantly better solutions again.

\section{Conclusion and Outlook}
\label{sec:conclusion}

We presented DB-BERT, a database tuning system that extracts tuning hints from text documents. Our experiments demonstrate that such hints lead to significantly better tuning results.

In future work, we will consider more diverse tuning objectives. Currently, DB-BERT is limited to optimizing metrics such as latency or throughput that can be easily measured. However, there are other, important metrics that are difficult to measure. For instance, many parameters (e.g., the \verb|fsync| parameter in Postgres) allow increasing performance if willing to accept a small risk of data loss. Database manuals typically contain warnings detailing such risks. We plan to extend DB-BERT to extract information on metrics that are difficult to measure from the manual. Thereby, it can support users in finding parameter settings that maximize performance while complying with constraints on other metrics.

\bibliographystyle{ACM-Reference-Format}
\bibliography{library}

\balance

\end{document}